\documentclass[12pt]{article}
\usepackage{amssymb,amsmath}
\usepackage{cancel}
\usepackage{palatino}
\usepackage{longtable}
\usepackage{color}
\numberwithin{equation}{section}
\newtheorem{theorem}{Theorem}[section]     
\newtheorem{definition}[theorem]{Definition}
\newtheorem{defi}[theorem]{Definition}

\newtheorem{rmk}[theorem]{Remark}
\newtheorem{corollary}[theorem]{Corollary}

\newtheorem{remark}[theorem]{Remark}

\def\d{\partial}

\def\f{\frac}

\def\proof{\noindent\hspace{2em}{\itshape Proof: }}
\def\QEDclosed{\mbox{\rule[0pt]{1.3ex}{1.3ex}}} % for a filled box

\def\QED{\QEDclosed} 
\def\endproof{\hspace*{\fill}~\QED\par\endtrivlist\unskip}
\newcommand{\eqa}{\begin{eqnarray}}
\newcommand{\eeqa}{\end{eqnarray}}
\newcommand{\beq}{\begin{equation}}
\newcommand{\eeq}{\end{equation}}

\begin{document}
\title{Flat $F$-manifolds, Miura invariants and integrable systems of conservation laws}
\author{Alessandro Arsie* and Paolo Lorenzoni**\\
\\
{\small *Department of Mathematics and Statistics}\\
{\small The University of Toledo,}
{\small 2801 W. Bancroft St., 43606 Toledo, OH, USA}\\
{\small **Dipartimento di Matematica e Applicazioni}\\
{\small Universit\`a di Milano-Bicocca,}
{\small Via Roberto Cozzi 53, I-20125 Milano, Italy}\\
{\small *alessandro.arsie@utoledo.edu,  **paolo.lorenzoni@unimib.it}}

\date{}

\maketitle

\begin{abstract}
We extend some of the results proved for scalar equations in \cite{ALM,ALM2},  to the case of systems of integrable conservation laws. In particular, for such systems we prove that the eigenvalues of a matrix obtained from the quasilinear part of the system are invariants under Miura transformations and we show how these invariants are related to dispersion relations. Furthermore, focusing on one-parameter families of dispersionless systems of integrable conservation laws associated to the Coxeter groups of rank $2$ found in \cite{ALcomplex}, we study the corresponding integrable deformations up to order $2$ in the deformation parameter $\epsilon$. Each family contains both bi-Hamiltonian and non-Hamiltonian systems of conservation laws and therefore we use it to probe to which extent the properties of the dispersionless limit impact the nature and the existence of integrable deformations. It turns out that a part two values of the parameter all deformations of order one in $\epsilon$ are Miura-trivial, while all those of order two in $\epsilon$ are essentially parameterized by two arbitrary functions of single variables (the Riemann invariants) both in the bi-Hamiltonian and in the non-Hamiltonian case.
In the two remaining cases, due to the existence of non-trivial first order deformations, there is an additional functional parameter.
%In the two remaining cases we prove the existence of non trivial first order deformations parametrized by an arbitrary function of a single variable. 
\end{abstract}

\newpage
\tableofcontents
\section{Introduction}

In this paper, we consider systems of PDEs in evolution form of the following type:
\begin{small}
\begin{equation}\label{intsys}
u^i_t=A^i_j({\bf u})u^j_x+\epsilon(B^i_j({\bf u})u^j_{xx}+B^i_{jk}({\bf u})u^j_xu^k_x)
+\epsilon^2(C^i_j({\bf u})u^j_{xxx}+C^i_{jk}({\bf u})u^j_{xx}u^k_x+C^i_{jkl}({\bf u})u^j_xu^k_xu^l_x)+\dots
\end{equation}
\end{small}
where $i=1,\dots n$ and $u^i(x,t)$ are functions of the space variable $x$ and the time variable $t$. Special cases of these systems include deformations of bi-Hamiltonian systems of hydrodynamic type and have been deeply investigated starting from the work of Dubrovin and Zhang \cite{DZ} (see also \cite{L,LZ2,DLZ,DLZaim,AL2,LZ3,CPS1,CPS2}), not only in relation to problems in the theory of integrable systems, but also with an essential focus to important aspects of the theory of Gromov-Witten invariants. 
In this work, our efforts are focused only on issues related to the classification and the integrability of systems \eqref{intsys}. 

To this aim, we say that the system \eqref{intsys} is integrable if
\begin{enumerate}
\item its dispersionless limit,  given by 
\begin{equation}\label{dlim}
u^i_t=A^i_j({\bf u})u^j_x,\quad i=1,...,n,
\end{equation}
is integrable in the sense of Tsarev, as a semi-Hamiltonian system \cite{Tsarev}. This implies the existence of a distinguished coordinate  system (defined up to reparameterizations) $\{r^1, \dots, r^n\}$. The coordinates $r^i, i=1,\dots, n$ are called Riemann invariants and in these coordinates the system \eqref{dlim} appears in diagonal form: 
 $$r^i_t=v^i(r^1,...,r^n)r^i_x,\quad i=1,...,n.$$
 The characteristic velocities of the system, namely the functions $v^i$ satisfy Tsarev's semi-Hamiltonian condition, i.e. 
 \begin{equation}\label{tscond}
 \d_k\left(\f{\d_jv^i}{v^j-v^i}\right)=\d_j\left(\f{\d_kv^i}{v^k-v^i}\right),\qquad i\ne j\ne k\ne i.
 \end{equation}
 This condition guarantees the existence of a family of symmetries and of densities of conservation laws, depending on $n$ arbitrary functions of a single variable. It turns out that the systems that are integrable according to Tsarev's definition, coincide with the systems of conservation laws that admit Riemann invariants (see \cite{Sevennec}).

\item  Every symmetry (commuting flow) of the dispersionless limit \eqref{dlim} can be extended to a symmetry of the complete system \eqref{intsys}. If this extension is obtained up to a certain order in $\epsilon$, then the system \eqref{intsys} is called integrable up to that order in $\epsilon$.
\end{enumerate}
The symmetries approach to integrability has a long history (see for instance \cite{MSY}, \cite{MSS}, \cite{SM}). In this paper following \cite{LZ} we consider symmetries represented by possibly infinite series. Our definition of integrability in the scalar case  coincides with the definition of \emph{formal integrability} given in \cite{LZ}.   

With this notion of integrability, one can think of systems \eqref{intsys} that are integrable as a sort of {\em integrable deformation} of dispersionless systems \eqref{dlim} that are Tsarev's integrable. This point of view has been widely adopted in the literature, see for instance \cite{DZ} (for the bi-Hamiltonian case) and \cite{DZ,ALM,ALM2} for the scalar case. In this paper, we adopt this point of view and we extend it to general systems (not necessarily Hamiltonian or bi-Hamiltonian) of the form \eqref{intsys}. 
A trivial way to obtain an integrable system of the form \eqref{intsys} is simply to start from a dispersionless system \eqref{dlim}, integrable in the Tsarev's sense, and apply to it a Miura transformation ${\bf u}\to{\bf w}$:
\beq\label{Miura}
w^i=\sum_{k=0}^{\infty} \epsilon^k F^i_k({\bf u},{\bf u}_x,...),\qquad {\rm deg}(F_k)=k,
\eeq
where it is assumed that $F_k$ are homogeneous differential polynomials of degree $k$ and that the leading term of the transformation is invertible. This last assumption implies the invertibility of the entire transformation.
%\begin{equation}\label{Miurat}
%u^i\to\tilde{u}^i=...
%\end{equation}
Due to the presence of this mechanism to construct integrable systems of the form \eqref{intsys} that are essentially trivial, one of the fundamental issues of the theory of integrable systems is to classify  systems of the form \eqref{intsys} only up to Miura transformations. Under some suitable assumptions (locality of the bi-Hamiltonian structure and semisemplicity) this problem has been solved completely in the bi-Hamiltonian case. In this case, it turns out that non trivial bi-Hamiltonian structures (which in the bi-Hamiltonian set-up control the deformations) are parameterized by $n$ arbitrary functions of one variable \cite{LZ2,DLZ}. These are called {\em central invariants} because of their invariance with respect to Miura transformations. For any choice of the dispersionless limit and of
 the central invariants there exists a unique (up to Miura transformations) bi-Hamiltonian structure  (see \cite{LZ3} and \cite{CPS1} for the scalar case and \cite{CPS2} for the general semisimple case). 

In the general case of systems of the form \eqref{intsys}, there are very few results available in the literature, and besides few exceptions they are all confined to the scalar case (see  \cite{DZ,D,ALM,ALM2}).
More specifically:
\begin{itemize}
\item In \cite{DZ} it has been proved that {\em every scalar} equation of the form \eqref{intsys} can be reduced to its dispersionless limit through a generalized Miura transformation (i.e. a Miura transformation in which the functions $F_k$ in \eqref{Miura} are not necessarily differential polynomials, but belong to a more general class of functions). 

A different proof of this result (obtained following the approach of \cite{BGI} based on the study of solutions of transport equations) is given in \cite{ALM2}. 
\item In \cite{ALM2} it has been proved that for {\em scalar} equations of the form \eqref{intsys} each coefficient of the quasilinear part behaves like a scalar under Miura transformations. In particular, in this case each coefficient is invariant with respect to the subgroup of Miura transformations of the form 
\begin{equation}\label{Miurat2}
u^i\to w^i=u^i+...
\end{equation}
i.e. Miura transformations where $F_0$ in \eqref{Miura} is the identity map.
\item In Table 1 (taken from \cite{ALM2}) we summarize some conjectures. They deal with the number of functional parameters that are needed to parameterize the (integrable) deformations in the scalar case. In each instance, it is possible to choose the functional parameters among the coefficients of the quasilinear part. 

%(in the bi-Hamiltonian case, it is necessary to consider the coefficient of $\epsilon^2$ of the first non-trivial associated flow). 

\begin{table}[h]
\begin{center}
\begin{tabular}{|l|c|l|}

    \hline
	{\bf Type of deformations} & {\bf Numbers of invariants} &
	  {\bf References}    \\ 
	\hline
General viscous deformations  & $2$ & \cite{LZ}  \\
\hline
General dispersive deformations  & $\infty$ & \cite{ALM2} and \cite{S}  \\
\hline
Viscous conservation laws  & $1$  & \cite{ALM}  \\
\hline
Dispersive conservation laws  & $\infty$ & \cite{ALM2}  \\
	\hline
Hamiltonian conservation laws  & $\infty$  & \cite{D,Dtalk}  \\
\hline
\end{tabular}
\end{center}
\caption{\footnotesize Summary of conjectures in the scalar case.}
\label{tab_results}
\end{table} 
\end{itemize} 

In this work, we will focus on the general case of systems, and in particular to the case of systems of conservation laws, i.e. we will deal with systems of the form 
\begin{equation}\label{intsysofCL}
u^i_t=\d_x\left[A^i({\bf u})+\epsilon(B^i_j({\bf u})u^j_{x}\right)
+\epsilon^2\left(C^i_j({\bf u})u^j_{xx}+C^i_{jk}({\bf u})u^j_{x}u^k_x) +\dots\right]
\end{equation}

In the first part of the paper, generalizing the results of \cite{ALM2} we show that the {\em eigenvalues} $\lambda_i({\bf u},p)$ of the matrix
 \begin{equation}\label{Miuramatrix1}
 M^i_j({\bf u}, p):=A^i_j({\bf u})+B^i_j({\bf u})p+C^i_j({\bf u})p^2+\dots
 \end{equation} associated to systems of the form \eqref{intsys} are scalar with respect to Miura transformations \eqref{Miura}. Consequently they are invariant with respect to the subgroup of Miura transformations preserving the dispersionless limit (i.e. Miura transformations \eqref{Miura} with $F_0$ being the identity map).  Let us remark that there is an important difference between the scalar case (studied in \cite{ALM2}) and the general case. In the first case the single coefficients of the quasilinear part are invariant, while in the latter case this is not true anymore. 
 
We call the matrix \eqref{Miuramatrix1} the \emph{Miura matrix} of the system and its eigenvalues $\lambda_i({\bf u}, p)$ \emph{Miura invariants} of the system. 
 
 Furthermore, we show how the Miura invariants are related to {\em dispersion relations} for systems of the form \eqref{intsys}. Indeed it turns out that dispersion relations are just given by $\omega_j(k)=-k\lambda_j({\bf u}_0, ik)$, where ${\bf u}_0\in \mathbb{R}^n$ is a constant solution of \eqref{intsys} (the need for choosing a constant solution comes from the fact that the dispersion relations arise when one linearizes the system \eqref{intsys} around a specified constant reference solution ${\bf u}_0$). 
 
In the second part of the paper, we focus on deformations of dispersionless integrable systems of conservation laws associated the to the families of bi-flat structures on the space of orbits of the Coxeter groups $B_2$ and $I_2(m)$ constructed in \cite{ALcomplex}. These families depend on one parameter $c$ (a scalar constant). For each choice of the constant one obtains a dispersionless integrable systems of conservation laws in two fields. The main reason to consider these examples is the fact that for generic values of $c$ one obtains a system of conservation laws which is not Hamiltonian (at least with respect to a local Hamiltonian structure). On the other hand, for a specific value of the parameter ($c=-\frac{3}{4}$ in the case of $B_2$ and $c=0$ in the case of $I_2(m)$), one gets a bi-Hamiltonian system (which is just the principal hierarchy of the Frobenius manifolds of $B_2$ and $I_2(m)$ respectively).
In other words, within the same family (parameterized by $c$) we have dispersionless integrable systems that are Hamiltonian (really bi-Hamiltonian) and non-Hamiltonian. This is extremely useful if one wants to probe to which extent the properties of the dispersionless limit impact the nature and the existence of (integrable) deformations. We restrict ourselves to deformations that preserve the form of conservation law for the PDEs under consideration. 

We have computed explicitly (using the formalism of \cite{AL}) all deformations up to the second order $\epsilon.$
In the case of $B_2$ the results obtained can be summarized as follows:
\begin{enumerate}
\item The integrable deformations have a branch for  $c=-\frac{3}{4}$, and surprisingly also for $c=-1$ and $c=-\frac{1}{2}$ (see below).  For the case $c=-\frac{3}{4}$, the dispersionless limit is bi-Hamiltonian, while for $c\neq -\frac{3}{4}$ the dispersionless limit is not Hamiltonian (at least with respect to a local structure). Furthermore, in the computation of the deformations, it is not possible to compute the deformation for general values of $c$ and then substitute $c=-\frac{3}{4}$, because in several quantities this produces infinities. In this sense, $c=-\frac{3}{4}$ is a singular limit of the family to be deformed. Moreover, from the geometric point of view, the dispersionless limit for $c=-\frac{3}{4}$ arises from a Frobenius manifold, while for $c\neq -\frac{3}{4}$ arises from a bi-flat $F$-manifold (see \cite{ALcomplex}).
\item  The deformations at order one in $\epsilon$ are Miura trivial, in the sense that they can be eliminated via  suitable Miura transformations except for the cases $c= -1, c= -\frac{1}{2}$. This pinpoints to the fact that {\em viscous} integrable systems (using a terminology taken from \cite{ALM}) might not be so abundant. For $c=-1$, the first order deformations can not be eliminated via a Miura transformation and they depend on a single arbitrary function of one variable (that turns out to be one of the Riemann invariants). Analogously, for $c=-\frac{1}{2}$, the first order deformations are not Miura trivial and they again depend on a single arbitrary function of one variable. 
\item The deformations at the order two in $\epsilon$ depend on two arbitrary functions of a single variable (the Riemann invariants), again for all values of $c$, except for $c=-1$ and $c=-\frac{1}{2}$. It is however, much more straightforward to see this in the Frobenius case where $c=-\frac{3}{4}.$ See also the important remark \ref{ImportantRemark}. For the cases $c=-1$ and $c=-\frac{1}{2}$ it happens that the second order deformations depend on three functional parameters of a single variable (due to the additional parameter appearing at the first order). 
\item The cases $c=-1$ and $c=-\frac{1}{2}$ turn out to be peculiar due to the fact that one of the primary flows  (and all the corresponding higher flows) of the principal hierarchy are degenerate.
  \end{enumerate}
  In the case of $I_2(m)$ we have analogue results (clearly for $m=4$ this is not sursprising since $B_2\sim I_2(4)$). For generic values of the parameter ($c\ne 0,\pm 2$) first order deformations are always trivial and second order deformations depend on two functional parameters. For $c=0$ (the bi-Hamiltonian case) we have similar results while for $c=\pm 2$ we have an additional parameter appearing at the first order.

% Despite the fact that these results are limited to a special class of examples and are confined to deformations of order up to $\epsilon^2$, they do provide some precise indications about what one should expect. First of all, in the case of systems the existence of integrable viscous deformations is not granted at all, unlike what happens in the scalar case (see \cite{ALM}). Secondly, while in the scalar case the coefficients of the integrable deformation at a certain order in $\epsilon$ are determined imposing commutativity of the flows at a order in $\epsilon$, which is higher than the order at which the undetermined coefficient appears, this "feedback" effect does not appear in the case of systems. Namely, all the coefficients at a certain order in $\epsilon$ are determined imposing the commutativity of the flows at the same order in $\epsilon$ (this at least up to order two, but notice that this does not happen even at the order two in $\epsilon$ in the scalar case, see \cite{ALM}).
 
The paper is organized as follows. In Section \ref{sec2} we introduce the Miura invariants and show how the are related to dispersion relations. In Section \ref{FlatbiFlatB2} we briefly recall the notions of flat and bi-flat $F$-manifolds, the principal hierarchy associated to these geometric structures and the examples related to Coxeter groups. In Section 4 we study in details the example of $B_2$. In subsection \ref{PH} we study the first flows of the principal hierarchy, in subsection \ref{sec4} we study the first order integrable deformations, while the second order deformations in $\epsilon$ are analyzed in subsection \ref{sec5}, together with some other important remarks. The  Section \ref{sec6} summarizes the results obtained for the integrable deformations of the principal hierarchy associated to $I_2(m)$ (dihedral groups), while the final Section \ref{sec7} provides some conclusions.

\section{Miura invariants}\label{sec2}
In this Section, we introduce a specific set of $0$-tensors, that we call {\em Miura invariants} associated to 
 any system of evolutionary PDEs of the form
\beq\label{system}
u^i_t=X^i({\bf u},{\bf u}_x,....),\qquad i=1,...,n,
\eeq
where 
$$X^i=A^i_j({\bf u})u^j_x+\epsilon(B^i_j({\bf u})u^j_{xx}+B^i_{jk}({\bf u})u^j_xu^k_x)
+\epsilon^2(C^i_j({\bf u})u^j_{xxx}+C^i_{jk}({\bf u})u^j_{xx}u^k_x+C^i_{jkl}({\bf u})u^j_xu^k_xu^l_x)+\dots$$
More specifically Miura invariants are invariants of the system \eqref{system} under the subgroup of Miura transformations \eqref{Miura} of the form:
\beq\label{subMiura}
w^i=u^i+\sum_{k=1}^{\infty} \epsilon^k F^i_k({\bf u},{\bf u}_x,...),\qquad {\rm deg}(F_k)=k.
\eeq
The Miura invariants defined in this note generalize the notion of invariants introduced by Dubrovin, Liu and Zhang in the context of bi-Hamiltonian evolutionary systems (see \cite{DLZ,LZ2,DLZaim}) and they also generalize one of the results of \cite{ALM2}, where it was proved that for any evolutionary {\em scalar} PDE of the form \eqref{system}, the coefficients of the quasilinear part of the PDE behave as scalars under Miura transformations \eqref{Miura} (so if $F_0$ is the identity map, then {\em each} coefficent of the quasilinear part is invariant).

 We denote by
\beq\label{Miurainv}
u^i=\sum_k \epsilon^k G^i_k({\bf w},{\bf w}_x,...),\qquad {\rm deg}(G_k)=k,
\eeq
the inverse of the Miura transformation \eqref{Miura}, where the leading part $F_0$ is assumed to be at least a local diffeomorphism. In the new ``coordinates''  $w's$ the system \eqref{system} reads
$$w^i_t=\left(\f{\d w^i}{\d u^j}+\f{\d w^i}{\d u^j_x}\d_x+\f{\d w^i}{\d u^j_{xx}}\d_x^2+\dots\right)X^j$$
where the right hand side must be rewritten in terms of ${\bf w},{\bf w}_x,...$.

We will now focus on the quasilinear part of the system \eqref{system}. This is by definition
\begin{eqnarray*}
X_{quasilinear}^i&=&A^i_j({\bf u})u^j_x+\epsilon B^i_j({\bf u})u^j_{xx}
+\epsilon^2C^i_j({\bf u})u^j_{xxx}+\dots=\\
&=&\left(A^i_j({\bf u})+\epsilon B^i_j({\bf u})\d_x
+\epsilon^2C^i_j({\bf u})\d_x^2+\dots\right)u^j_x
\end{eqnarray*}
In particular, let us consider the Miura matrix $M^i_j({\bf u},p):=A^i_j({\bf u})+B^i_j({\bf u})p+C^i_j({\bf u})p^2+\dots$ constructed with the matrix coefficients of the quasilinear part. 
\begin{definition}
We call {\em Miura invariants} the eigenvalues $\lambda_i({\bf u},p)$ of the matrix $M^i_j({\bf u},p)$ defined above.
\end{definition}

To justify this definition we will study  the behaviour of $\lambda_i$ under Miura transformation. 
The key idea consists in writing the quasilinear part of \eqref{system} as a linear operator acting on ${\bf u}_x$ and studying as its symbol behaves under Miura transformations. A similar idea was used in \cite{DLZaim} to prove the invariance of central invariants.

\begin{theorem}\label{mainthm}
Consider the Miura transformation \eqref{Miura} and its inverse \eqref{Miurainv}. The Miura invariants of \eqref{system} transform as scalars, namely $\tilde \lambda_i({\bf w})=\lambda_i(G_0({\bf w}))$, where ${\bf u}=G_0({\bf w})$ and $G_0$ is the (local) inverse of $F_0$. In particular the Miura invariants are invariant under Miura transformations with $F_0=\rm{Id}$.
\end{theorem}
\proof

As a preliminary observation, we consider the system
\beq\label{trivsys}
u^i_t=u^i_x,\qquad i=1,...,n.
\eeq
The Miura matrix in this case is the identity matrix. It is easy to check that the system \eqref{trivsys}
after a Miura transformations maintains its form. Indeed
$$w^i_t=\left(\f{\d w^i}{\d u^j}+\f{\d w^i}{\d u^j_x}\d_x+\f{\d w^i}{\d u^j_{xx}}\d_x^2+\dots\right)u^j_x=w^i_x.$$
In particular this means that the quasilinear part and the corresponding Miura matrix do not change in this case.

Now we analyze what is the effect of a Miura transformation on the quasilinear part of \eqref{system}. Due to the fact that $u^i$ is related to $w^i$ via \eqref{subMiura}, the only part of the operator
$$\left(\f{\d w^i}{\d u^j}+\f{\d w^i}{\d u^j_x}\d_x+\f{\d w^i}{\d u^j_{xx}}\d_x^2+\dots\right)$$
that can affect the quasilinear part is given by
$$\left(\f{\d F_0^i}{\d u^j}+\f{\d F_1^i}{\d u^j_x}\d_x+\f{\d F_2^i}{\d u^j_{xx}}\d_x^2+\dots\right).$$
Similarly the only part of $u^i_x$ that can provide a contribution to the quasilinear part is
$$\left(\f{\d G_0^i}{\d w^j}+\f{\d G_1^i}{\d w^j_x}\d_x+\f{\d G_2^i}{\d w^j_{xx}}\d_x^2+\dots\right)w^j_x.$$
In other words, the transformed quasilinear part can be obtained taking the quasilinear part of
\begin{small}
\beq\label{trrop}
\left(\f{\d F_0^i}{\d u^l}+\epsilon\f{\d F_1^i}{\d u^l_x}\d_x+\dots\right)
\left(A^l_m({\bf u})+\epsilon B^l_m({\bf u})\d_x+\dots\right)\left(\f{\d G_0^m}{\d w^j}+\epsilon\f{\d G_1^m}{\d w^j_x}\d_x+\dots\right)w^j_x
\eeq
\end{small}
evaluated at ${\bf u}=G_0({\bf w})$. This means that the Miura matrix $M^i_j({\bf u},p)$ transforms according to the following rule involving only the symbols of the operators appearing in \eqref{trrop}:
 \beq\label{trr}
 \tilde{M}^i_j({\bf w},p)=\left(\f{\d F_0^i}{\d u^l}+\f{\d F_1^i}{\d u^l_x}p+\dots\right)
M^l_m({\bf u},p)\left(\f{\d G_0^m}{\d w^j}+\f{\d G_1^m}{\d w^j_x}p+\dots\right),
\eeq
where, in the right hand side, ${\bf u}=G_0({\bf w})$. Notice that the invariance of the system \eqref{trivsys} implies
\beq\label{strr}
\left(\f{\d F_0^i}{\d u^l}+\f{\d F_1^i}{\d u^l_x}p+\dots\right)
\left(\f{\d G_0^l}{\d w^j}+\f{\d G_1^l}{\d w^j_x}p+\dots\right)=\delta^i_j.
\eeq
Applying the transformation rule \eqref{trr} to the Miura matrix of the system
$$u^i_t=X^i({\bf u},{\bf u}_x,....)-\lambda u^i_x,\qquad i=1,...,n,$$
and taking into account \eqref{strr}, we obtain
$$\tilde{M}^i_j({\bf w},p)-\lambda\delta^i_j=\left(\f{\d F_0^i}{\d u^j}+\f{\d F_1^i}{\d u^j_x}p+\dots\right)
(M^j_m({\bf u},p)-\lambda\delta^j_m)\left(\f{\d G_0^m}{\d w^l}+\f{\d G_1^m}{\d w^l_x}p+\dots\right).$$ 
The above formula immediately implies that the eigenvalues of the matrix $M^i_j({\bf u},p)$ transform as scalars:
$$\tilde{\lambda}_i({\bf w})=\lambda_i(G_0({\bf w})).$$
 In particular they are invariant under Miura transformations of the form \eqref{subMiura}.
% $$w^i=u^i+\sum_{k>0} \epsilon^k F^i_k({\bf u},{\bf u}_x,...),\qquad {\rm deg}(F_k)=k.$$
\endproof

\begin{rmk}
Notice that for $n>1$ not only the coefficients of the quasilinear part (namely the matrices $A^i_j({\bf u}), B^i_j({\bf u}), C^i_j({\bf u})$, etc.) are not separately invariants, but even their eigenvalues are not preserved under Miura transformations. 
\end{rmk}

Recall that an evolutionary system like \eqref{system} is called Miura trivial if there exists a Miura transformation that reduces it to the form $w^i_t=A^i_j({\bf w})w^j_x$. Using Theorem \ref{mainthm}, we have immediately the following:
\begin{corollary}
A necessary condition for a system of the form \eqref{system} to be Miura trivial is that its Miura invariants $\lambda_i({\bf u},p)$ do not depend on $p$.
\end{corollary}
In general, this condition is not sufficient. In \cite{ALM2} it has been conjectured that in the scalar integrable case this condition becomes also sufficient, and more in general, that two integrable scalar equation are related by a Miura transformation of the form \eqref{subMiura} if and only if they have the same Miura invariants. The results of the present paper suggest a similar scenario also in the  non-scalar case.

\begin{rmk}
In the scalar case, Miura invariants coincide with the coefficients of the quasilinear part of the equation (see \cite{ALM2}).
%$$\lambda= A(u)+B(u)p+C(u)p^2+\dots$$
\end{rmk}

\begin{remark}
Miura invariants can be defined in the same way also in the non semisimple case.
\end{remark} 

\subsection{Miura invariants and dispersion relations}
In this section we show how the notion of Miura invariants is related to the  more familiar notion of dispersive relation.
   
Any linear constant coefficients scalar PDE of evolutionary type on unbounded domains ($x\in \mathbb{R}$) admits traveling wave solutions of the form $u(x,t)=\exp(ikx-i\omega t)$, whenever the frequency $\omega=\omega(k)$ satisfy  a particular relation with respect to the wave number $k$. This relation is called {\em dispersive relation} and in general there are $m$ different dispersive relations if the highest order of the time derivative appearing in the PDE is $m$. 

%In our case we will restrict to evolutionary PDE (and system) with only one time derivative, so the dispersive relation is really a unique function characteristic of the PDE under analysis. 

Of course nonlinear PDEs in general do not admit solutions of the form $u(x,t)=\exp(ikx-\omega t).$ However, it is still possible to linearize a nonlinear PDE near a particular solution and analyze the behavior of the linear approximation. 
For instance, linearizing the KdV equation $u_t=uu_x+\gamma u_{xxx}$ near the constant solution $u_0=1$ with $u(x,t)=u_0+\epsilon v(x,t)$, we obtain the linear PDE $v_t=v_x+\gamma v_{xxx}.$
This readily admits a traveling wave solution of the form $v(x,t)=\exp(ikx-i\omega t)$ where the dispersion relation in this case is $\omega(k)=\gamma k^3-k$. Since $\omega$ is a nonlinear function of $k$ we say that the behavior of the KdV equation is dispersive (at least near the linearization). Instead we say that the behavior is non-dispersive if $\omega$ is a linear function of $k$. 

More in general if we deal with a scalar linear PDE with constant real coefficients, even-numbered spatial derivative terms are diffusive while odd-numbered spatial derivative terms are dispersive, in general (provided that there are odd-numbered spatial derivative of order greater than one). However, the situation is much more complicated for the case of systems, even for those that are linear with constant coefficients. 

%In this case, indeed, the behavior of each term depends not only on the order of the derivative but also on the eigenvalues of each coefficient matrix. 

The case of systems  \eqref{system} (setting $\epsilon=1$):
\begin{equation}\label{system2} u^i_t=A^i_j({\bf u})u^j_x+(B^i_j({\bf u})u^j_{xx}+B^i_{jk}({\bf u})u^j_xu^k_x)
+(C^i_j({\bf u})u^j_{xxx}+C^i_{jk}({\bf u})u^j_{xx}u^k_x+C^i_{jkl}({\bf u})u^j_xu^k_xu^l_x)+\dots
\end{equation}
can be treated in a similar way. Linearizing the system around a constant solution ${\bf u}={\bf c}\in \mathbb{R}^n$ we obtain a system of the form
\begin{equation}\label{linsysconstcoeff.eq}{\bf u}_t=\sum_{j} A_j {\bf u}_{(j)},\end{equation}
with  $(A_1)^i_j=A({\bf c})^i_j,\, (A_2)^i_j=B({\bf c})^i_j,\, (A_3)^i_j=C({\bf c})^i_j, $ and so on. 
 If we plug it in \eqref{linsysconstcoeff.eq}, we obtain that the vector ${\bf v}$ has to satisfy the equation 
\begin{equation}\label{eq1aux}-i\omega(k) {\bf v}=ik\sum_{j} (ik)^{j-1} A_j {\bf v}.\end{equation}
From equation \eqref{eq1aux} we see that  ${\bf v}$ is an eigenvector of the matrix 
\begin{equation}\label{matrixM}M:=\sum_{j} (ik)^{j-1} A_j.\end{equation}
 If we call $\lambda(k)$ the corresponding eigenvalue, then from equation \eqref{eq1aux} we obtain the dispersion relation 
$$\omega(k) =-k\lambda(k).$$
In general, there will be $n$ different dispersion relations ($\{\omega_j(k)=-k\lambda_j(k), \, j=1,\dots, n\}$), one for each eigenvalue of the matrix $M$. 
But the eigenvalues of $M$  coincide with the Miura invariants of \eqref{system2}  computed at ${\bf u}={\bf c}$ and $p=ik$. 

\section{Flat and bi-flat $F$-manifolds and Coxeter groups}\label{FlatbiFlatB2}
In this Section, we briefly recall the geometric structures that encode the existence of a dispersionless integrable hierarchy (not necessarily Hamiltonian), called {\em the principal hierarchy}.

\subsection{Flat $F$-manifold}

\begin{defi}
An $F$-manifold with compatible flat structure (or flat $F$-manifold)  $(M,\circ,\nabla,e)$ is 
 a manifold  equipped with a product $\circ : TM \times TM \rightarrow TM$ on the tangent spaces, a connection $\nabla$
 and a distinguished vector field $e$ such that
\begin{itemize}
\item the one parameter family of  connections
$$\nabla-\lambda\circ$$
is  flat and torsionless for any $\lambda$.
\item $e$ is the unit of the product and it is flat: $\nabla e=0$.
\end{itemize}
\end{defi}

Let $\Gamma^k_{ij}$ be the Christoffel symbols of $\nabla$ 
and $c^k_{ij}$ the structure constants of the product, then the fact that $\Gamma^k_{ij}-\lambda c^k_{ij}$ is flat and torsionless for
 any $\lambda$ is equivalent to the following conditions:
\begin{enumerate}
\item the connection $\nabla$ is torsionless and the product  $\circ$ is commutative,
\item the connection $\nabla$ is flat and the product $\circ$ is associative,
\item the tensor field $\nabla_l c^k_{ij}$ is symmetric in the lower indices.
\end{enumerate}

From condition 1,2,3 it follows that, in flat coordinates for $\nabla$ we have
$$c^i_{jk}=\d_j\d_k A^i.$$
The vector potential $A^i$ satisfies the associativity equations:
\begin{eqnarray*}
\d_j\d_l A^i\d_k\d_mA^l&=&\d_k\d_l A^i\d_k\d_mA^l\\
\end{eqnarray*}

\subsection{The principal hierarchy}
Given a flat $F$-manifold one can define an integrable hierarchy of quasilinear systems of evolutionary PDEs of the form
$$u_{t_{(p,l)}}=X_{(p,l)}\circ u_x,\qquad p=1,...,n\qquad l=0,1,2,3,...$$
The vector fields $X_{(p,l)}$ defining the hierarchy as coefficents of the formal expansion of flat sections of the deformed flat connection:
$$(\nabla-\lambda \circ)\left(X_{(p,0)}+X_{(p,1)}\lambda+X_{(p,2)}\lambda^2+...\right)=0.$$
\newline
This means that the vector fields $X_{(p,0)}$ are flat and the remaining ones are obtained by means of the recursive relations
\beq\label{recrel}
\nabla X_{(p,l+1)}=X_{(p,l)}\circ.
\eeq
In flat coordinates $(v^1,...,v^n)$ the flows of the hierarchy are systems of conservation laws:
$$v_{t_{(p,l)}}=X_{(p,l)}\circ v_x=\d_x X_{(p,l+1)}.$$
If the product is semisimple, then the canonical coordinates $(r_1,...,r_n)$ (where $c^i_{jk}=\delta^i_j\delta^i_k$) are Riemann invariants 
 of this system.

If $\nabla$ is the Levi-Civita connection of $\eta$ and $\eta$ is invariant w.r.t the product: 
$$<X\circ Y,Z>=<X,Y\circ Z>,\qquad\forall X,Y,Z,$$
where $<\cdot,\cdot>$ is the bilinear form defined by $\eta$, the bi-flat $F$-manifolds becomes Frobenius manifolds. Indeed:
\begin{itemize}
\item 
$\eta_{il}A^l=\d_i F$. In other words oriented associativity equations become WDVV associativity equations:
\begin{eqnarray*}
\d_j\d_h\d_i F\eta^{il}\d_l\d_k\d_m F&=&\d_j\d_k \d_iF\eta^{il}\d_l\d_h\d_m F\\
\d_n\d_i\d_j F&=&\eta_{ij}\\
\end{eqnarray*}
\item the principal hierarchy becomes Hamiltonian w.r.t. the Dubrovin-Novikov bracket associated with $\eta$. In flat coordinates
 $$\eta_{ij}X^j_{(p,l)}=\d_i h_{(p,l)}$$
and the flows of the principal hierarchies can be written as
$$v_{t_{(p,l)}}=X_{(p,l)}\circ v_x=\d_x X_{(p,l+1)}=P\delta  H_{(p,l+1)}$$
where $H[v]=\int  h_{(p,l+1)}(v)\, dx$ and $P^{ij}=\eta^{ij}\d_x$.
\end{itemize}

To conclude this subsection we recall the definition of bi-flat $F$-manifolds.

\begin{defi}
A \emph{bi-flat}  $F$-manifold is a manifold equipped with two different flat structures $(\nabla,\circ,e)$ and  $(\nabla^{*},*,E)$
  related by the following conditions
\begin{enumerate}
\item $X*Y=E^{-1}\circ X\circ Y,\quad\forall X,Y$.
\item $[e,E]=e$,
\item ${\rm Lie}_E \circ=\circ$,
\item $(d_{\nabla}-d_{\nabla^{*}})(X\,\circ)=0,\quad\forall X,$ where $d_{\nabla}$ is the exterior covariant derivative.
\end{enumerate}
\end{defi}

Whenever one has an underlying bi-flat $F$ manifold structure, it is possible to introduce a recursive scheme for the principal hierarchy which is different from \eqref{recrel} and which generalizes the so-called Lenard-Magri chains. These recurrence relations are called \emph{twisted Lenard-Magri chains} and were developed in \cite{ALimrn}. More precisely, on any (semisimple) bi-flat F-manifold, the following recursive scheme (generalizing the classical bi-Hamiltonian recurrence relations of Lenard-Magri) is available:  
$$d_{\nabla^{(2)}}\left(E\circ X_{(p,\alpha)}\right)=d_{\nabla^{(1)}}X_{(p+1,\alpha)},$$
where $(X_{(0,1)},\dots,X_{(0,n)})$ is a frame of flat
 vector fields. The corresponding equations of the associated hierarchy are in this case given by:
\beq\label{thy}
u^i_{t_{(p,\alpha)}}= [d_{\nabla^{(2)}}\left(E\circ X_{(p-1,\alpha)}\right)]^i_j\,u^j_x=      (d_{\nabla^{(1)}}X_{(p,\alpha)})^i_j\,u^j_x,\qquad i=1,\dots,n,\,p=1,2,3,\dots.
\eeq

\subsection{Bi-flat $F$-manifolds and Coxeter groups}
Let $G$ be a Coxeter group and $|\mathcal{H}/G|$ the number of orbits for the action of $G$ on the collection of reflecting hyperplanes. In \cite{ALcomplex} it has been conjectured that the orbit space of each finite Coxeter group is equipped with a natural bi-flat $F$-manifold structure that depends on $|\mathcal{H}/G|-1$ parameters. The conjecture is supposed to hold under the assumption that the flat coordinates of $\nabla$ are basic {\em polynomial} invariants and that the dual product has the special form
$$*=\f{1}{N}\sum_{H\in \mathcal{H}}\frac{d\alpha_H}{\alpha_H}\otimes\pi_H,$$
where $\alpha_H$ is a linear form defining the mirror $H$, $\mathcal{H}$
 is  the collection of the reflecting hyperplanes, $\pi_H$ denotes the orthogonal projection onto the orthogonal complement (with respect to a suitable bilinear form) of the hyperplane $H$ and $N$ is a normalizing factor. 
This conjecture  has been verified for Weyl groups of rank $2$, $3$, and $4$ and for the groups $I_2(m)$. In the first case one has one parameter in the cases  $B_2,B_3,B_4,F_4$ and no parameter in the remaining cases. In the second case we have one parameter in the even case and no parameter in the odd case. 

Below we write some examples of one-parameter families of vector potentials. 

\subsubsection{$B_2$}
\begin{eqnarray*}
A_{B_2}^1&=& -\f{2}{3}\left(c+\f{3}{4}\right)u_1^3+u_1u_2,\\
A_{B_2}^2&=& -\f{1}{6}(c+1)(2c+1)u_1^4+\f{1}{2}u_2^2.
\end{eqnarray*}
The value of the parameter corresponding to the Frobenius manifold structure is  $c=-\f{3}{4}$.

\subsubsection{$B_3$}
\begin{eqnarray*}
A^1_{B_3}&=&\f{1}{3}\left(c+\f{4}{3}\right)\left(c+\f{5}{4}\right)u_1^4-\left(c+\f{5}{4}\right)u_2u_1^2
+u_1u_3+\f{3}{8}u_2^2,\\
A^2_{B_3}&=&\f{4}{45}\left(c+\f{5}{4}\right)(c+2)(c+1)u_1^5-\f{1}{9}(c+1)u_2u_1^3\\
&&-\f{1}{2}(c+1)u_2^2u_1+u_2u_3,\\
A^3_{B_3}&=&\f{1}{2}u_3^2-\f{1}{135}\left(c+\f{3}{2}\right)(c+1)(8c^2+20c+13)u_1^6+\\
&&\f{1}{3}\left(c+\f{5}{4}\right)(c+1)\left(c+\f{3}{2}\right)u_2u_1^4-\f{1}{2}(c+1)\left(c+\f{3}{2}\right)u_2^2u_1^2\\
&&+\f{1}{8}\left(c+\f{3}{2}\right)u_2^3.
\end{eqnarray*}
The value of the parameter corresponding to the Frobenius manifold structure is  $c=-\f{5}{4}$.

\subsubsection{$B_4$}
\begin{eqnarray*}
A^1_{B_4}&=&\left(-\frac{32}{5}c^3-\frac{131}{15}c^2-\frac{953}{240}c-\frac{77}{128}\right)u_1^5+8\left(c+\frac{7}{16}\right)\left(c+\frac{11}{24}\right)u_1^3u_2+\left(-\frac{8}{3}c-\frac{7}{6}\right)u_3u_1^2\\
&&+\left(-2c-\frac{7}{8}\right)u_2^2u_1+u_4u_1+\frac{2}{3}u_3u_2,\\
A^2_{B_4}&=&\left(-\frac{64}{15}c^4-\frac{392}{45}c^3-\frac{1163}{180}c^2-\frac{499}{240}c-\frac{63}{256}\right)u_1^6+\frac{4}{3}\left(c+\frac{7}{16}\right)\left(c+\frac{3}{8}\right)u_2u_1^4\\
&&+\left(-\frac{2}{9}c-\frac{1}{12}\right)u_3u_1^3+4\left(c+\frac{3}{8}\right)^2u_1^2u_2^2+\left(-\frac{8}{3}c-1\right)u_3u_2u_1+u_4u_2+\frac{4}{9}u_3^2\\
&&+\left(-\frac{2}{3}c-\frac{1}{4}\right)u_2^3,\\
A^3_{B_4}&=&\left(\frac{64}{7}c^5+\frac{440}{21}c^4+\frac{229}{12}c^3+\frac{1453}{168}c^2+\frac{3499}{1792}c+\frac{45}{256}\right)u_1^7\\
&&-\frac{96}{5}\left(c+\frac{5}{8}\right)\left(c^2+\frac{41}{48}c+\frac{47}{256}\right)\left(c+\frac{3}{8}\right)u_2u_1^5+\frac{1}{3}\left(c+\frac{3}{8}\right)\left(c+\frac{7}{16}\right)u_3u_1^4\\
&&+12\left(c+\frac{5}{8}\right)\left(c+\frac{3}{8}\right)\left(c+\frac{5}{12}\right)u_1^3u_2^2+\left(-\frac{1}{2}c-\frac{3}{16}\right)u_3u_2u_1^2\\
&&+\left(-2c^2-2c-\frac{15}{32}\right)u_1u_2^3+\left(-\frac{4}{3}c-\frac{1}{2}\right)u_3^2u_1+\frac{1}{8}u_3u_2^2+u_4u_3,\\
A^4_{B_4}&=&\left(-\frac{128}{7}c^6-48c^5-\frac{158}{3}c^4-\frac{371}{12}c^3-\frac{983}{96}c^2-\frac{1393}{768}c-\frac{481}{3584}\right)u_1^8\\
&&+\frac{128}{3}\left(c+\frac{1}{2}\right)\left(c+\frac{3}{8}\right)\left(c+\frac{7}{16}\right)\left(c^2+\frac{41}{48}c+\frac{73}{384}\right)u_1^6u_2\\
&&-\frac{256}{15}\left(c+\frac{1}{2}\right)\left(c+\frac{3}{8}\right)\left(c+\frac{11}{24}\right)\left(c+\frac{7}{16}\right)u_3u_1^5\\
&&-32\left(c+\frac{1}{2}\right)\left(c+\frac{3}{8}\right)\left(c^2+\frac{5}{6}c+\frac{17}{96}\right)u_1^4u_2^2\\
&&+\frac{64}{3}\left(c+\frac{1}{2}\right)\left(c+\frac{3}{8}\right)\left(c+\frac{7}{16}\right)u_1^3u_2u_3+8\left(c+\frac{1}{2}\right)\left(c+\frac{3}{8}\right)^2u_1^2u_2^3\\
&&-\frac{32}{9}\left(c+\frac{1}{2}\right)\left(c+\frac{3}{8}\right)u_3^2u_1^2+\left(-\frac{16}{3}c^2-\frac{14}{3}c-1\right)u_3u_2^2u_1\\
&&+\left(-\frac{1}{3}c^2-\frac{1}{4}c-\frac{1}{24}\right)u_2^4+\left(\frac{8}{9}c+\frac{4}{9}\right)u_3^2u_2+\frac{1}{2}u_4^2.
\end{eqnarray*}
The vector potential for $B_4$ comes from a Frobenius potential for $c=-\f{7}{16}$.  

\subsubsection{$I_2$(m)} 
%For the case of $I_2(m)$ (dihedral groups) with $m$ even (if $m$ is odd, there is no parameter and the corresponding structure is always Frobenius):
\begin{eqnarray*}
A^1_{I_2(m)}&=& u_1u_2-\frac{2c}{\frac{m}{2}+1}u_1^{\frac{m}{2}+1} ,\\
A^2_{I_2(m)}& =&\frac{1}{2}u_2^2+\left(-\frac{m}{(4m-4)}c^2+\frac{m}{m-1} \right) u_1^m 
\end{eqnarray*}
The vector potential for $I_{2}(m)$ comes from a Frobenius potential for $c=0$. 

\begin{rmk}
If $m$ is odd the above formulas still define a solution of the oriented associativity equations for any value of $c$. However, in this case the flat coordinates are no longer polynomials. Therefore, the case with $m$ odd does not provide a counteraxample to the conjecture mentioned above. Since
 the requirement that the flat coordinates are polynomials  does not play any role in the present context, we will not restrict to the even case in the rest of the paper.
\end{rmk}

Based on what we have presented in this Section, it is clear that to each Coxeter group there is associated a family of integrable dispersionless hierarchies, whenever the corresponding vector potential depends on parameters.  If the vector potential does not depend on parameters, then there is associated a unique integrable dispersionless hierarchy: the principal hierarchy of the Frobenius manifold associated with the Coxeter group. In the case of families, there is a special value of the parameter for which the associated integrable dispersionless hierarchy is the principal hierarchy originating from the Frobenius manifold associated  with the Coxeter group. For generic values of the parameter the bi-flat structure does not come from a Frobenius manifold structure and  the principal hierarchy is not Hamiltonian (at least w.r.t. a local Hamiltonian structure).  In the next Section, we will study in detail the case of $B_2$ and of its integrable deformations.
 
\section{The case of $B_2$}
Starting from the results of \cite{ALcomplex}, we provide the principal hierarchy associated to the orbit space of $B_2$ and we analyze its integrable deformations. This is a particularly significant example because of the fact that the underlying geometric structures (bi-flat $F$-manifold) for the case of $B_2$ appear in a family depending on a parameter. 

\subsection{The principal hierarchy for $B_2$}\label{PH}
In this paper we focus on a one-parameter family of bi-flat structures associated with the Coxeter group $B_2$. The primary flows of the hierarchy are 
\begin{eqnarray}
\label{B2-1}
u_{t_{1,0}}&=&\d_x\left[v-2\left(c+\f{3}{4}\right)u^2\right]\\
\label{B2-2}
v_{t_{1,0}}&=&\d_x\left[-\f{4}{3}\left(c+\f{1}{2}\right)(c+1)u^3\right]
\end{eqnarray}
and
\begin{eqnarray}
\label{tr1}
u_{t_{2,0}}&=&u_x,\\
\label{tr2}
v_{t_{2,0}}&=&v_x.
\end{eqnarray}
Starting from the primary flows one can reconstruct all the flows of the principal hierarchy by means of the recursive relations \eqref{recrel}. For instance, the first
flows obtained from \eqref{B2-1},\eqref{B2-2} and \eqref{tr1} and \eqref{tr2} are
\begin{eqnarray}
\label{fl11-1}
u_{t_{1,1}}&=&\d_x\left[-2\left(c+\f{3}{4}\right)u^2v+\left(\f{5}{3}c^2+\f{5}{2}c+\f{23}{24}\right)u^4+\f{v^2}{2}\right]\\
\label{fl11-2}
v_{t_{1,1}}&=&\d_x\left[\f{8}{5}\left(c+\f{3}{4}\right)\left(c+\f{1}{2}\right)(c+1)u^5-\f{4}{3}\left(c+\f{1}{2}\right)(c+1)u^3v\right]
\end{eqnarray}
and
\begin{eqnarray}
\label{fl21-1}
u_{t_{2,1}}&=&\d_x\left[-\f{4}{3}\left(c+\f{3}{4}\right)u^3+uv\right]\\
\label{fl21-2}
v_{t_{2,1}}&=&\d_x\left[-\f{1}{4}\left(c+\f{1}{2}\right)(c+1)u^4+\f{v^2}{2}\right]
\end{eqnarray}
respectively.

In the case $c=-\f{3}{4}$ all the flows of the principal hierarchy  can be written in the bi-hamiltonian form. This is the case corresponding to Frobenius manifold of $B_2$. In the remaining cases we have systems of conservation laws admitting Riemann invariants,
 that is semi-Hamiltonian systems. A special set of Riemann invariants are the canonical coordinates of the product which are
 $$r_1=-cu^2-u^2+v,\qquad r_2=-cu^2-\f{1}{2}u^2+v.$$
\subsection{The general strategy}
Here we outline the general strategy to compute non-trivial integrable deformations of the principal hierarchy associated with a flat $F$-manifolds. The steps are as follows:
\newline
\newline
{\bf Step 1}. We reduce the general deformation at some order in $\epsilon$ (for our case only order one and two) for the selected flow of the principal hierarchy to the simplest possible form using Miura transformations. 
% In some cases this is done after having imposed the integrability (namely the commutativity with a selected symmetry). 
\newline
\newline
{\bf Step 2}. We choose a symmetry and we impose commutativity between the selected deformed flow and the selected symmetry (which is to be deformed at the same order in $\epsilon$). The commutativity is also imposed up to the chosen order in $\epsilon$ (for the cases examined here, only order one and two). 
\newline
\newline
{\bf Step 3}. We check that {\em all} the symmetries (and not just the selected one) of the system (\ref{B2-1}, \ref{B2-2}) can be extended to the relevant order in the deformation parameter.
\newline
\newline
{\bf Step 4}. We check that all deformed the symmetries pairwise commute up to the relevant order in $\epsilon.$

\begin{remark}
In some cases, the order of the first and second steps are switched.
\end{remark}

\begin{remark}
Let us underline that in all the examples we dealt with, Step 3 and Step 4 follow automatically once Step 2 is completed, in the sense that no new conditions are necessary for Step 3 and Step 4 to be fulfilled. This means that in all the examples analyzed, extending one symmetry is enough to extend all of them (to the same order) and automatically implies that all the extended symmetries commute among each other. This is consistent with the conjecture formulated in \cite{F}, where it is stated that for a system with $n$ components, it is enough to prolong $n$ symmetries (in our examples, it seems that even the prolongation of one symmetry does  the job). 
\end{remark}

In general,  since we are dealing with systems of conservation laws, the commutativity between the flow $u_t=\d_x(\alpha_1),v_t=\d_x(\alpha_2)$
 and the flow $u_\tau=\d_x(\beta_1),v_\tau=\d_x(\beta_2)$ is equivalent to the involutivity of the $1$-forms defining
the flows with respect to the Poisson bracket \cite{AL}:
\beq\label{truePoisson1}
\{\alpha, \beta\}_i:=(\d_x^{s+1}\beta_k)\f{\d\alpha_i}{\d u^k_{(s)}}-(\d_x^{s+1}\alpha_k)\f{\d \beta_i}{\d u^k_{(s)}}. 
\eeq

The formula \eqref{truePoisson1} is just a special case of a more general formula that reduces to \eqref{truePoisson1} in a special set of coordinates. 
 Indeed, consider two PDEs of the form 
 \begin{eqnarray}\label{hameqne}
u^i_t&=&P^{ij}\alpha_j=\left(g^{ij}\d_x-g^{il}\Gamma_{lk}^ju^k_x\right)
\alpha_j,\qquad i=1,\dots,n,\\
u^i_{\tau}&=&P^{ij}\beta_j=\left(g^{ij}\d_x-g^{il}\Gamma_{lk}^ju^k_x\right)
\beta_j,\qquad i=1,\dots,n,
 \end{eqnarray}
were the $1$-forms $\alpha$ and $\beta$ are not necessarily closed (they are not necessarily the variational derivative of some functionals) and $g^{ij}$ is a flat contravariant metric with corresponding Christoffel symbols $\Gamma_{lk}^j$. 
Then the two flows defined by these two PDEs commute if and only if 
 the Poisson bracket defined by
\beq\begin{split}\label{generalPoisson}
\{\alpha, \beta\}_i=\partial_x^s\left(g^{kl}\partial_x\beta_l+\Gamma^{kl}_mu^m_x\beta_l\right)\f{\partial\alpha_i}{\partial u^k_{(s)}}-\partial_x^s\left(g^{kl}\partial\alpha_l+\Gamma^{kl}_mu^m_x\alpha_l\right)\f{\partial\beta_i}{\partial u^k_{(s)}}\\
+\left(\alpha_k\partial_x\beta_l-\beta_k\partial_x\alpha_l\right)\Gamma^{lk}_i-\alpha_k\beta_l\left[\Gamma^{k}_{is}\Gamma^{sl}_m-\Gamma^l_{is}\Gamma^{sk}_m\right]u^m_x,
\end{split}\eeq
vanishes. For instance, all the flows of the principal hierarchy associated with a flat $F$-manifold $(M,\nabla,\circ,e)$ can be written in the form  \eqref{hameqne} where  $g$ is any (pseudo)-metric satisfying the condition $\nabla g=0$ and the 1-forms $\alpha$ are obtained from the vector fields
 $X$ defining the principal hierarchy lowering the index with $g$: $\alpha_i=g_{ij}X^j$. In a given system of flat coordinates without loss of generality
 we can choose $g_{ij}=\delta_{ij}$. In this case  it is immediate to check that the \eqref{generalPoisson} reduces to \eqref{truePoisson1}.
For the development of the full theory and further details, see \cite{AL}.

\subsection{Deformations at the first order in $\epsilon$}\label{sec4}
In this Section, we study the (integrable) deformations at the first order in $\epsilon$ for the dispersionless systems associated to $B_2$ constructed in Section \ref{FlatbiFlatB2}. We do not spell out in more detail the general strategy outlined in the previous subsection for this case, since we are going to provide more details for the second order deformations, where the computations are considerably more involved.

 As a first step we have to select a flow in the principal hierarchy.  For $c\ne -1,-\f{1}{2}$ we choose  the flow $(1,0)$, corresponding to  the system (\ref{B2-1}, \ref{B2-2}). Since for $c=-\f{1}{2}$ and for $c=-1$ this system is degenerate, for these values of $c$ we choose the flow  $(2,1)$. This is given by
\begin{eqnarray*}
u_{t_{2,1}}&=&\d_x\left[uv-\f{1}{3}u^3\right]\\
v_{t_{2,1}}&=&\d_x\left[\f{1}{2}v^2\right]\\
\end{eqnarray*}
in the case $c=-\f{1}{2}$ and by
\begin{eqnarray*}
u_{t_{2,1}}&=&\d_x\left[uv+\f{1}{3}u^3\right]\\
v_{t_{2,1}}&=&\d_x\left[\f{1}{2}v^2\right]\\
\end{eqnarray*}
in the case $c=-1$.

To impose integrability, we need to choose a symmetry (see Step 2 above). In the case $c\ne-\f{1}{2},-1$ we choose as symmetry the flow $(2,1)$ (\ref{fl21-1},\ref{fl21-2}) of the principal hierarchy, while in the case $c=-\f{1}{2}$ e $c=-1$ we choose the flow $(2,2)$. In the first case we have the system
\begin{eqnarray*}
u_{t_{2,2}}&=&\d_x\left[-\f{1}{3}u^3v+\f{1}{2}uv^2+\f{1}{15}u^5\right],\\
v_{t_{2,2}}&=&\d_x\left[\f{1}{6}v^3\right],\\
\end{eqnarray*}
while in the second case we get the system
\begin{eqnarray*}
u_{t_{2,2}}&=&\d_x\left[\f{1}{3}u^3v+\f{1}{2}uv^2+\f{1}{15}u^5\right],\\
v_{t_{2,2}}&=&\d_x\left[\f{1}{6}v^3\right].\\
\end{eqnarray*}

\subsubsection{The case $c=-\frac{3}{4}$}
For the case $c=-\frac{3}{4}$ we first apply the reducing Miura transformation to the {\em general} deformation (i.e. without imposing integrability conditions). In doing so, in particular trying to cancel the entire deformation using a Miura transformation, one obtains the integrability conditions {\em automatically}, without imposing them. Thus the integrability conditions at this level appear as {\em necessary} conditions for the first order deformation to be Miura trivial. 

For $c=-\frac{3}{4}$, the general first order deformation of the system is given by:
\begin{eqnarray*}
u_t &=&\partial_x\left(v+\epsilon(a_{11}u_x+a_{12}v_x) \right)=\partial_x(\omega_{1,0}+\epsilon\omega_{1,1}) ,\\
 v_t& =& \partial_x\left(\frac{1}{12}u^3+\epsilon(a_{21}u_x+a_{22}v_x) \right)=\partial_x(\omega_{2,0}+\epsilon\omega_{2,1})
\end{eqnarray*}
where $a_{11}, \dots a_{22}$ are arbitrary (analytic) functions of $u,v$.
Now we consider Miura transformations of the form $u^i \mapsto w^i=u^i +\epsilon\partial_x M_i$ for $i=1,2$ (so $u^1=u$ and $u^2=v$), where $M_i$ are differential polynomials of degree zero in the dependent variables. Via a straightforward calculation we see that in order for the first order deformation to be Miura trivial, it is necessary and sufficient that the following systems of PDEs is satisfied (we write it using the old variables $u,v$):
\begin{eqnarray*}
a_{11}+\frac{1}{4}u^2\partial_{v}M_1-\partial_{u}M_2&=&0,\\
a_{12}+\partial_{u}M_1-\partial_{v}M_2&=&0,\\
a_{21}+\frac{1}{4}u^2\partial_{v}M_2-\frac{1}{4}u^2\partial_{u}M_1&=&0,\\
a_{22}+\partial_{u}M_2-\frac{1}{4}u^2\partial_{v}M_1&=&0.
\end{eqnarray*}
These four PDEs are just the coefficients of the homogeneous part of degree one of the forms $\omega_{1,0}+\epsilon\omega_{1,1}$ and $\omega_{2,0}+\epsilon\omega_{2,1}$ after the application of the Miura transformation. In particular, the first and the second PDEs are the coefficients in front of $u_{x}$ and $v_x$ respectively in the homogeneous part of degree one  of the form $\omega_{1,0}+\epsilon\omega_{1,1}$ after Miura transformation, while the third and the fourth PDEs are the coefficients in front of $u_x$ and $v_x$ respectively in the in the homogeneous part of degree one  of the form $\omega_{2,0}+\epsilon\omega_{2,1}$ after Miura transformation.  
Observe that for the first and last equation of this system to be satisfied, it is necessary that $a_{11}=-a_{22}$. Analogously for the second and third equation to be fulfilled, it is necessary that $a_{21}=-\frac{1}{4}u^2 a_{12}$. These two conditions turn out to be a posteriori exactly the {\em integrability conditions} of involutivity with the symmetries. 
Imposing these necessary conditions for the system about to be satisfied, we have that it is reduced to the following (two of the equations become redundant):
\begin{eqnarray}\label{syseq1cFrob}
a_{11}+\frac{1}{4}u^2\partial_{v}M_1-\partial_{u}M_2&=&0,\\\label{syseq2cFrob}
a_{12}+\partial_{u}M_1-\partial_{v}M_2&=&0.
\end{eqnarray}
While in general it is difficult to find an explicit solution of this system, one can use general theorems to prove that a solution of the does exist. The main theorem in this context is the so-called Cauchy-Kowalevski theorem which we now state in a form useful for our goals:
\begin{theorem}\label{CKThm}
Let ${\bf M}: \mathbb{R}^n \rightarrow \mathbb{R}^n$ and consider the following problem:
\begin{equation}\label{CK}\partial_{u^n}{\bf M}=\sum_{i=1}^{n-1}A_i({\bf u},{\bf M})\partial_{u^i}{\bf M}+B({\bf u},{\bf M}), \quad {\bf M}(u^1,\dots, u^{n-1},0)=\Phi(u^1, \dots, u^{n-1}),\end{equation}
where $B, A_1, \dots, A_{n-1}, \Phi$ are analytic maps, each analytic in a neighborhood of the origin of their respective domains. Then there exists a neighborhood of the origin in $\mathbb{R}^n$ on which there exists a unique analytic function ${\bf M}:\mathbb{R}^n \rightarrow \mathbb{R}^n$ solving the Cauchy problem \eqref{CK}.
\end{theorem}

To apply this Theorem, we rewrite the system \eqref{syseq1cFrob}, \eqref{syseq2cFrob} as (where $u$ corresponds to the first variable and $v$ to the second):
$$\partial_1 {\bf M}=A\partial_2{\bf M}+{\bf b},$$
where 
$${\bf M}=\begin{bmatrix}
  M_1     \\
 M_2
\end{bmatrix},\quad 
A=\begin{bmatrix}
  0      & 1 \\
    \frac{1}{4}u^2      & 0
\end{bmatrix},\quad {\bf b}=\begin{bmatrix}
  -a_{12}     \\
   a_{11}
\end{bmatrix},$$
and $a_{11},a_{12}$ are known but unspecified functions of $u,v$.
Then, assuming that $a_{11}$ and $a_{12}$ are analytic in a neighborhood of the origin in $\mathbb{R}^2$, the Cauchy-Kowalevski theorem guarantees that there exists a unique analytic solution to the above system in a neighborhood of the origin with ${\bf M}(0, v)=0$. Of course, one can choose a different initial condition for ${\bf M}$ (as long as it is analytic). 

\subsubsection{The case $c\neq -\frac{3}{4},\, -1, \,  -\frac{1}{2}$}
Here we study the Miura triviality of the first order deformations for the generic case of $c$ arbitrary. Besides excluding the case $c=-\frac{3}{4}$, it is necessary to exclude also the cases $c=-1, c=-\frac{1}{2}$, because in this case it is the starting point of the principal hierarchy that is degenerate, and one needs to deal with these two cases separately.  

Furthermore, in this case,  {in order to simplify the computations we first impose integrability and then we look for the reducing Miura transformation.   One can easily verify \emph{a posteriori} that also in this case  the integrability conditions coincide with the compatibility
  of the system of PDEs governing the reducing Miura transformation.
 
% compatibility conditions are precisely the compatibility conditions among the system of PDEs. However, in this case, it is much more difficult to find out the compatibility conditions (i.e. the integrability conditions) examining directly the PDEs involved. 
%
%we do not consider arbitrary first order deformations, but directly integrable first order deformations (i.e. the integrability constraints have been already imposed). This is because if one considers initially general first order deformations and one tries to find Miura transformations that reduce as much as possible the deformations, the PDE involved are incompatible, since the compatibility conditions are precisely the compatibility conditions among the system of PDEs. However, in this case, it is much more difficult to find out the compatibility conditions (i.e. the integrability conditions) examining directly the PDEs involved. 

Thefore, after having imposed the integrability conditions, the first order deformations are given by: 
\begin{eqnarray*}
u_t &=&\partial_x\left[v-2\left(c+\f{3}{4}\right)u^2+\epsilon \alpha \right]=\partial_x(\omega_{1,0}+\epsilon\omega_{1,1}) ,\\
 v_t& =& \partial_x\left[-\f{4}{3}\left(c+\f{1}{2}\right)(c+1)u^3+\epsilon \beta \right]=\partial_x(\omega_{2,0}+\epsilon\omega_{2,1})
\end{eqnarray*}
where 
$$\alpha=\frac{1}{16}\frac{\left(-16u^2(c+1)((c+\frac{3}{4})u^2-\frac{1}{2}v)(c+\frac{1}{2})a_{12}-2b_{21}\right)u_{x}}{u(c+\frac{3}{4})((c+\frac{3}{4})u^2-\frac{1}{2}v)}+a_{12}v_x,$$
$$\beta=-\frac{2b_{21}u_x}{(4c+3)u^2-2v}+\frac{1}{16}\frac{\left(16u^2(c+1)((c+\frac{3}{4})u^2-\frac{1}{2}v)(c+\frac{1}{2})a_{12}+2b_{21}\right)v_{x}}{u(c+\frac{3}{4})((c+\frac{3}{4})u^2-\frac{1}{2}v)}.$$
Here the functions $a_{12}$ and $b_{21}$ are arbitrary functions that control the deformations (those that remains after imposing integrability).

As usual, we apply Miura transformations of the form $u^i \mapsto w^i=u^i +\epsilon\partial_x M_i$ for $i=1,2$ (so $u^1=u$ and $u^2=v$), where $M_i$ are differential polynomials of degree zero in the dependent variables. Via a straightforward calculation we see that in order for the first order deformation to be Miura trivial, it is necessary and sufficient that the a system of four PDEs is satisfied (we write it using the old variables $u,v$). Of these PDEs, the first and the last are the same, so we write only the first three PDEs:

\begin{eqnarray*}
-\frac{2(2c^2+3c+1)a_{12}u}{4c+3}-\frac{2b_{21}}{u(4c+3)(4cu^2+3u^2-2v)}+\\
-\partial_{u}M_2-(4u^2c^2+6u^2c+2u^2)\partial_{v}M_1&=&0,\\
a_{12}+(4uc+3u)\partial_{v}M_1-\partial_{v}M_2+\partial_{u}M_1&=&0,\\
-\frac{2b_{21}}{4cu^2+3u^2-2v}+(4u^2c^2+6u^2c+2u^2)\partial_{u}M_1+\\
-(4uc+3u)\partial_{u}M_2-(4u^2c^2+6u^2c+2u^2)\partial_{v}M_2 &=&0.
\end{eqnarray*}
These four PDEs (the fourth is omitted since it is equal to the first up to a sign) are just the coefficients of the homogeneous part of degree one of the forms $\omega_{1,0}+\epsilon\omega_{1,1}$ and $\omega_{2,0}+\epsilon\omega_{2,1}$ after the application of the Miura transformation. In particular, the first and the second PDEs are the coefficients in front of $u_{x}$ and $v_x$ respectively in the homogeneous part of degree one of the form $\omega_{1,0}+\epsilon\omega_{1,1}$ after Miura transformation, while the third and the (missing) fourth PDEs are the coefficients in front of $u_x$ and $v_x$ respectively in the homogeneous part of degree one  of the form $\omega_{2,0}+\epsilon\omega_{2,1}$ after Miura transformation.

Solving for $\partial_{u}M_2$ in the first equation and for $\partial_{v}M_2$ in the second equation and substituting the resulting expressions in the third PDE, one gets that the third equation is automatically satisfied. So the system of PDEs that control the existence of a Miura transformation eliminating all the first order deformations is given by the first two equations and can be written as
$$\partial_{u} {\bf M}=A\partial_{v}{\bf M}+{\bf b},$$
where 
\begin{small}
$${\bf M}=\begin{bmatrix}
  M_1     \\
 M_2
\end{bmatrix},\, A=\begin{bmatrix}
  -4\left(c+\f{3}{4}\right)u     & 1 \\
   -4u^2(c+1)\left(c+\f{1}{2}\right)    & 0
\end{bmatrix},\, {\bf b}=\begin{bmatrix}
  -a_{12}     \\
   -\frac{(c+1)\left(c+\f{1}{2}\right)a_{12}u}{c+\f{3}{4}}-\frac{b_{21}}{8u\left(c+\f{3}{4}\right)\left(\left(c+\f{3}{4}\right)u^2-\f{v}{2}\right)}
\end{bmatrix}.$$
\end{small}
Therefore, assuming that $a_{12}$ and $b_{21}$ are analytic functions in a suitable neighborhood (not of the origin in $\mathbb{R}^2$ this time, since the vector valued function ${\bf b}$ will not be analytic there in general), we can still invoke Theorem \ref{CKThm} to claim the existence of a local solution ${\bf M}$.

\begin{remark}
Although the Miura transformations eliminating the first order deformations for $c\ne -1,-\f{1}{2}$ were identified using only one particular flow of the principal hierarchy, it turns out that they automatically reduce to the dispersionless limit the entire principal hierarchy, if they do so for one of the flows.
\end{remark}

\subsubsection{Case $c=-1$}
The cases with $c=-1$ and $c=-\frac{1}{2}$ present a substantial departure from the previous cases. Indeed in these cases, the first order deformations are {\em not} Miura trivial as we shall see in a moment. 

For the case $c=-1$, we first impose the integrability conditions and we find that the first order deformations are given by 

\begin{eqnarray*}
u_t &=&\partial_x\left[uv+\frac{1}{3}u^3+\epsilon(a_{12}v_{x}+a_{11}u_{x})\right]=\partial_x(\omega_{1,0}+\epsilon\omega_{1,1}) ,\\
 v_t& =& \partial_x\left[\frac{1}{2}v^2+\epsilon\left(F\left(\frac{1}{2}u^2+v\right)u^2u_{x}+F\left(\frac{1}{2}u^2+v\right)uv_{x}-a_{11}uu_{x}-a_{11}v_{x}\right) \right]=\\
  & =&\partial_x(\omega_{2,0}+\epsilon\omega_{2,1}).
\end{eqnarray*}
where $a_{12}, a_{11}$ are arbitrary functions of $u,v$ and $F$ is an arbitrary function of the single variable $\frac{1}{2}u^2+v$, which is one of the Riemann invariants. 
Applying Miura transformations of the usual form, via a straightforward calculation we see that in order for the first order deformation to be Miura trivial, it is necessary and sufficient that the following system of four PDEs is satisfied (we write it using the old variables $u,v$ and we omit one of the equations since two of them are essentially the same equation):
\begin{eqnarray*}
a_{11}-u\partial_{u}M_2&=&0,\\
a_{12}-u\partial_{v}M_2-u^2\partial_{v}M_1+u\partial_{u}M_1&=&0,\\
F\left( \frac{1}{2}u^2+v\right)u-a_{11}+u\partial_{u}M_2&=&0.
\end{eqnarray*}
These three PDEs are obtained in the following way. The first two are the coefficients of $u_x$ and $v_x$ respectively in the homogeneous part of degree one of $\omega_{1,0}+\epsilon\omega_{1,1}$, after the Miura transformation has been performed. The last equation is the coefficient in front of $v_x$ in the homogeneous part of degree one of $\omega_{2,0}+\epsilon\omega_{2,1}$, after the Miura transformation has been performed. The coefficient in front of $u_x$ in the homogeneous part of degree one of $\omega_{2,0}+\epsilon\omega_{2,1}$ after the Miura transformation is the same as the third equation multiplied by $u$.

The system is clearly incompatible (since the first and the third equation can not be simultaneously satisfied), which means that the function $F$ can not be eliminated from the first order deformation. On the other hand it is possible to obtain an explicit solution of the system formed by the first and the second equation (which means we can eliminate $a_{11}$ and $a_{12}$ from the deformation). In this way, we obtain the non-trivial first order deformations: 
\begin{eqnarray*}
u_t &=&\partial_x\left[uv+\frac{1}{3}u^3\right] ,\\
 v_t& =& \partial_x\left[\frac{1}{2}v^2+\epsilon\left(F\left(\frac{1}{2}u^2+v\right)u^2u_{x}+F\left(\frac{1}{2}u^2+v\right)uv_{x}\right) \right]
\end{eqnarray*}

\subsubsection{Case $c=-\frac{1}{2}$}
For the case $c=-\frac{1}{2}$, again we first impose the integrability conditions and we find that the first order deformations are given by 
\begin{eqnarray*}
u_t &=&\partial_x\left[uv-\frac{1}{3}u^3+\epsilon\left(F\left(-\frac{1}{2}u^2+v\right)uu_{x}-a_{22}u_{x}+a_{12}v_{x}\right)\right] =\\
  & = &\partial_x(\omega_{2,0}+\epsilon\omega_{2,1}),\\
 v_t& =& \partial_x\left[\frac{1}{2}v^2+\epsilon\left(a_{22}v_{x}-a_{22}uu_{x}\right)\right]=\partial_x(\omega_{2,0}+\epsilon\omega_{2,1}),
\end{eqnarray*}
where $a_{12}, a_{22}$ are arbitrary functions of $u,v$ and $F$ is an arbitrary function of the single variable $-\frac{1}{2}u^2+v$, which is the other Riemann invariant. 
Applying Miura transformations of the usual form, via a straightforward calculation we see that in order for the first order deformation to be Miura trivial, it is necessary and sufficient that the following system of four PDEs is satisfied (we write it using the old variables $u,v$ and we omit one of the equations since two of them are essentially the same equation):
\begin{eqnarray*}
F\left(-\frac{1}{2}u^2+v\right)u-a_{22}-u\partial_{u}M_2&=&0,\\
a_{12}-u\partial_{v}M_2+u^2\partial_{v}M_1+u\partial_{u}M_1&=&0,\\
a_{22}+u\partial_{u}M_2&=&0.
\end{eqnarray*}
These three PDEs are obtained in the following way. The first two are the coefficients of $u_x$ and $v_x$ respectively in the homogeneous part of degree one of  the form $\omega_{1,0}+\epsilon\omega_{1,1}$, after the Miura transformation has been performed. The last equation is the coefficient in front of $v_x$ in the homogeneous part of degree one of the form $\omega_{2,0}+\epsilon\omega_{2,1}$, after the Miura transformation has been performed. The coefficient in front of $u_x$ in the homogeneous part of degree one of  the form $\omega_{2,0}+\epsilon\omega_{2,1}$ after the Miura transformation is the same as the third equation multiplied by $-u$.

The system is clearly incompatible (since the first and the third equation can not be simultaneously satisfied), which means that the function $F$ can not be eliminated from the first order deformation. On the other hand it is possible to obtain an explicit solution of the system formed by the second and the third equation. In this way, we obtain the non-trivial first order deformations: 
\begin{eqnarray*}
u_t &=&\partial_x\left[uv-\frac{1}{3}u^3+\epsilon \,F\left(-\frac{1}{2}u^2+v\right)uu_{x}\right],\\
 v_t& =&  \partial_x\left[\frac{1}{2}v^2\right].
\end{eqnarray*}

We therefore conclude this Section with the following result, the proof of which stems from what we have seen above:
\begin{theorem}
The integrable first order deformations of the principal hierarchy of $B_2$ are Miura trivial for $c\neq -1, -\frac{1}{2}$ (assuming that the deformations are parametererized by analytic functions). Instead, the integrable first order deformations for $c= -1, -\frac{1}{2}$ are parameterized by an arbitrary function of a single variable (one of the Riemann invariants). 
\end{theorem}

\subsection{Deformations at the second order in $\epsilon$}\label{sec5}
In this section we study the second order non-trivial deformations of the principal hierarchy of $B_2$.

Here we spell out in slightly more detail the strategy used to compute the second order non-trivial integrable deformations of the principal hierarchy for $B_2$.
\newline
\newline
{\bf Step 1}. Same as the Step 1 above.}
\newline
\newline
{\bf Step 2}. We impose commutativity between the selected deformed flow and the selected deformed symmetry up to the second order in $\epsilon$. In the case $c\ne-\f{1}{2},-1$ due to the results of subection \ref{sec4} we do not have any first order correction 
 while in the cases $c=-\f{1}{2}$ and $c=-1$ we have the first order corrections obtained at the previous step.
 \newline
 \newline
{\bf Step 3}. We check that all the symmetries of the system (\ref{B2-1}, \ref{B2-2}) can be extended to the second order in the deformation parameter $\epsilon$. For $c\ne -1,-\f{1}{2}$ the symmetries are
\begin{eqnarray*}
u_{\tau}&=&-\f{1}{2(c+1)(2c+1)}\d_x\left(\f{1}{u^2}\f{\d h}{\d u}\right)\\
v_{\tau}&=&\d_x\left(\f{\d h}{\d v}\right)\\
\end{eqnarray*}
where the function $h$ is a solution of  
\beq\label{sym1}
u\f{\d^2 h}{\d u^2}+(4c+3)u^2\f{\d^2 h}{\d u\d v}+2(c+1)(2c+1)u^3\f{\d^2 h}{\d v^2}-2\f{\d h}{\d u}=0.
\eeq
For $c=-\f{1}{2}$ the symmetries are
\begin{eqnarray*}
u_{\tau}&=&\d_x\left(f(u,v)\right)\\
v_{\tau}&=&\d_x\left(g(v)\right)\\
\end{eqnarray*}
where the function $f$ and $g$ satisfies
\beq\label{sym2}
\f{\d f}{\d u}+u\f{\d f}{\d v}-g'(v)=0.
\eeq
Finally for $c=-1$ the symmetries are
\begin{eqnarray*}
u_{\tau}&=&\d_x\left(f(u,v)\right)\\
v_{\tau}&=&\d_x\left(g(v)\right)\\
\end{eqnarray*}
where the function $f$ and $g$ satisfies
\beq\label{sym3}
\f{\d f}{\d u}-u\f{\d f}{\d v}-g'(v)=0.
\eeq
It is known from the general theory \cite{Tsarev} that the general solutions of the equations (\ref{sym1},\ref{sym2},\ref{sym3}) depends on two arbitrary functions of single variables (the Riemann invariants). For special choices of these functions we get the flows of the principal hierarchy.
\newline
\newline
{\bf Step 4}. Same as Step 4 above.

\subsubsection{The case $c=-\f{3}{4}$}
\begin{theorem}
For $c=-\f{3}{4}$ the integrable second order deformations of the system  (\ref{B2-1},\ref{B2-2}) can be reduced to the following form
\begin{eqnarray}
\label{intdef1Fr}
u_{t_{(1,1)}}=\d_x\alpha_1&=&\d_x\left[v+\epsilon^2(a_{15}u_x^2+a_{16}v_x^2+a_{17} u_xv_x)\right]\\
\label{intdef2Fr}
v_{t_{(1,1)}}=\d_x\alpha_2&=&\d_x\left[\f{1}{12}u^3+\epsilon^2(a_{23}u_{xx}+a_{24}v_{xx}+a_{25}u_x^2-4\f{a_{25}}{u^2}v_x^2)\right]
\end{eqnarray}
where
\begin{eqnarray*}
a_{15}&=&-\f{3}{4}\f{\d^2 F}{\d u\d v},\\
a_{16}&=&\f{1}{u^2}\f{\d^2 F}{\d u\d v},\\
a_{17}&=&-\f{1}{2}\f{\d^2 F}{\d v^2},\\
a_{23}&=&\f{\d F}{\d u},\\
a_{24}&=&\f{\d F}{\d v},\\
a_{25}&=&\f{u^2}{8}\f{\d^2 F}{\d v^2}-\f{1}{4u}\f{\d F}{\d u}.
\end{eqnarray*}
and $F$ is the sum of two arbitrary functions of the Riemann invariants: 
$$F=F_1(\f{1}{4}u^2+v)+F_2(-\f{1}{4}u^2+v).$$
\end{theorem}

\emph{Proof}.
\newline
\newline
{\bf Step 1}. Following the general strategy we apply Miura  transformation of the form
\begin{eqnarray*}
\tilde{u}&=&u+\epsilon^2\d_x(\beta_{11}(u,v)u_x+\beta_{12}(u,v)v_x)+\mathcal{O}(\epsilon^2)\\
\tilde{v}&=&u+\epsilon^2\d_x(\beta_{21}(u,v)u_x+\beta_{22}(u,v)v_x)+\mathcal{O}(\epsilon^2)\\
\end{eqnarray*}
to general second order deformation of the system (\ref{B2-1},\ref{B2-2})
\begin{eqnarray*}
u_{t_{(1,0)}}=\d_x\alpha_1&=&\d_x\left[v+\epsilon^2(a_{13}u_{xx}+a_{14}v_{xx}+a_{15}u_x^2+a_{16}v_x^2+a_{17} u_xv_x)\right]\\
v_{t_{(1,0)}}=\d_x\alpha_2&=&\d_x\left[\f{1}{12}u^3+\epsilon^2(a_{23}u_{xx}+a_{24}v_{xx}+a_{25}u_x^2+a_{26}v_x^2+a_{27} u_xv_x)\right].
\end{eqnarray*}
It is easy to check that the terms in $\alpha_1$ containing $u_{xx}$ and $v_{xx}$ and the term in $\alpha_2$ containing $u_xv_x$ can be eliminated and the coefficent of $v_{x}^2$ in $\alpha_2$ can be chosen equal to the coefficient of $u_x^2$ multiply by $\f{-4}{u^2}$. 
\newline
\newline
{\bf Step 2}. Imposing commutativity up to the order $\epsilon^2$ with the flow 
\begin{eqnarray*}
u_{t_{(2,1)}}=\d_x\beta_1&=&\d_x\left[uv+\epsilon^2(b_{13}u_{xx}+b_{14}v_{xx}+b_{15}u_x^2+b_{16}v_x^2+b_{17} u_xv_x)\right]\\
v_{t_{(2,1)}}=\d_x\beta_2&=&\d_x\left[\f{1}{16}u^4+\f{1}{2}v^2+\epsilon^2(b_{23}u_{xx}+b_{24}v_{xx}+b_{25}u_x^2+b_{26}v_x^2+b_{27} u_xv_x)\right].
\end{eqnarray*}
we obatin all the unknown coefficients in terms of $a_{23}$ and $a_{24}$. These two remaining functions
  must satisfy the conditions
\begin{eqnarray*}
&& \f{\d a_{23}}{\d v}-\f{\d a_{24}}{\d u}=0\\
&&u^3\f{\d a_{24}}{\d v}-4u \f{\d a_{23}}{\d u}+4a_{23}=0
\end{eqnarray*}
The first condition tell us  that locally $a_{23}= \f{\d F}{\d u}$ and $a_{24}= \f{\d F}{\d v}$. Substituting in the  second condition we get
 the equation
$$4u\f{\d^2 F}{\d u^2} -u^3\f{\d^2 F}{\d v^2}-4\f{\d F}{\d u}=0,$$
whose general solution is $F=F_1(\f{1}{4}u^2+v)+F_2(-\f{1}{4}u^2+v)$.
\newline
\newline
{\bf Step 3}. We impose the commutativity between the  second order deformation of the system (\ref{B2-1},\ref{B2-2}) and the deformation of the general symmetry 
\begin{eqnarray*}
u_{\tau}=\d_x\beta_1&=&\d_x\left[\f{4}{u^2}\f{\d h}{\d u}+\epsilon^2(c_{13}u_{xx}+c_{14}v_{xx}+c_{15}u_x^2+c_{16}v_x^2+c_{17} u_xv_x)\right]\\
v_{\tau}=\d_x\beta_2&=&\d_x\left[\f{\d h}{\d u}+\epsilon^2(c_{23}u_{xx}+c_{24}v_{xx}+c_{25}u_x^2+c_{26}v_x^2+c_{27} u_xv_x)\right],
\end{eqnarray*}
where $h$ is assumed to satisfy the equation \eqref{sym1} with $c=-\f{3}{4}$. It turns out that the coefficients of the deformed symmetry are uniquely determined as functions of $F$ and $h$ and their partial derivatives:
\begin{footnotesize}
\begin{eqnarray*}
c_{13}&=&\f{h_{vvv}F_{v}u^2+4F_{u}h_{uvv}}{u},
\end{eqnarray*}
\begin{eqnarray*}
c_{14}&=&\f{4h_{vvv}F_{u}+h_{uvv}F_{v}}{u},
\end{eqnarray*}
\begin{eqnarray*}
c_{15}&=&-\f{1}{4}\f{u^3(F_{vv}h_{uvv}+F_{uv}h_{vvv}-2F_{u}h_{vvvv}-2F_{v}h_{uvvv})-6h_{vvv}F_{v}u^2+12h_{uv}F_{uv}-20F_{u}h_{uvv}}{u^2},
\end{eqnarray*}
\begin{eqnarray*}
c_{16}&=&-\f{u^3(F_{vv}h_{uvv}+F_{uv}h_{vvv}-2F_{v}h_{uvvv}-2F_{u}h_{vvvv})+4F_{u}h_{uvv}-4h_{uv}F_{uv}}{u^4},
\end{eqnarray*}
\begin{eqnarray*}
c_{17}&=&-\f{1}{2}\f{u^3(h_{vvv}F_{vv}-2F_{v}h_{vvvv})+4u(h_{uvv}F_{uv}-8F_{u}h_{uvvv})+4F_{vv}h_{uv}+4h_{vvv}F_{u}-8h_{uvv}F_{v}}{u^2},
\end{eqnarray*}
\begin{eqnarray*}
c_{23}&=&\f{u^3(F_{u}h_{vvv}+F_{v}h_{uvv})+4F_{u}h_{uv}}{u^2},
\end{eqnarray*}
\begin{eqnarray*}
c_{24}&=&h_{vvv}F_{v}u^3+4F_{u}h_{uvv}u+4F_{v}h_{uv},
\end{eqnarray*}
\begin{eqnarray*}
c_{25}&=&-\f{u^6(h_{vvv}F_{vv}-2F_{v}h_{vvvv})+4u^4(h_{uvv}F_{uv}-2F_{u}h_{uvvv})-4u^3(2F_{vv}h_{uv}+F_{u}h_{vvv}+2F_{v}h_{uvv})+16F_{u}h_{uv}}{16u^3},
\end{eqnarray*}
\begin{eqnarray*}
c_{26}&=&-\f{1}{4}\f{u^6(h_{vvv}F_{vv}-2F_{v}h_{vvvv})+4u^4(h_{uvv}F_{uv}-2F_{u}h_{uvvv})+4u^3(2F_{vv}h_{uv}+F_{u}h_{vvv}-2F_{v}h_{uvv})-16F_{u}h_{uv}}{u^5},
\end{eqnarray*}
\begin{eqnarray*}
c_{27}&=&-\f{1}{2}\f{u^3(F_{vv}h_{uvv}+F_{uv}h_{vvv}-2F_{v}h_{uvvv}-2F_{u}h_{vvvv})-2h_{vvv}F_{v}u^2-4F_{u}h_{uvv}}{u^2}.
\end{eqnarray*}
\end{footnotesize}
\newline
\newline
{\bf Step 4}. We verify that for any pair of solutions $(h_1,h_2)$ of \eqref{sym1} the corresponding deformed symmetries commute. This requires to use
 the differential consequences of the equations for $F$, $h_1$ and $h_2$. 
 
 \subsubsection{The case $c=-\f{1}{2}$}
\begin{theorem}
For $c=-\f{1}{2}$ the integrable deformations of the system (\ref{fl21-1},\ref{fl21-2}) can be reduced to the following form
\begin{eqnarray*}
\label{intdef1Fr}
u_{t_{(1,1)}}&=&\d_x\alpha_1=\d_x\left[uv-\f{1}{3}u^3+\epsilon(a_{11}u_x)+\epsilon^2(a_{16}v_x^2+a_{17}u_xv_x)\right]\\
\label{intdef2Fr}
v_{t_{(1,1)}}&=&\d_x\alpha_2=\d_x\left[\f{1}{2}v^2+\epsilon^2(a_{23}u_{xx}+a_{24}v_{xx}+a_{25}u_x^2+a_{26}v_x^2-a_{26}u_xv_x)\right]
\end{eqnarray*}
where
\begin{eqnarray*}
a_{11}&=&f(u,v)\\
a_{16}&=&g_{vv}+\f{1}{2}\f{ff_{vv}-h_v}{u}-\f{9}{2}\f{g_v}{u^2}+\f{2ff_v+h}{u^3}+\f{g}{u^4}-\f{2}{3}\f{f^2}{u^5}\\
a_{17}&=&-ug_{vv}-\f{1}{2}ff_{vv}+\f{4g_v}{u}-\f{ff_v}{u^2}+\f{2}{3}\f{f^2}{u^4}\\
a_{23}&=&g(u,v)\\
a_{24}&=&h(u,v)\\
a_{25}&=&-ug_v+\f{3g}{u}-\f{2}{3}\f{f^2}{u^2}\\
a_{26}&=&ug_{vv}-\f{6g_v}{u}+\f{8}{3}\f{ff_v}{u^2}+\f{3g}{u^3}-\f{4}{3}\f{f^2}{u^4}\\
%a_{27}&=&-u^2g_{vv}+6g_v-\f{8}{3}\f{ff_v}{u}-\f{3g}{u^2}+\f{4}{3}\f{f^2}{u^3}
\end{eqnarray*}
and
\begin{eqnarray*}
f&=& F(-\f{1}{2}u^2+v)u\\
g&=& G(-\f{1}{2}u^2+v)u^3+\f{f^2}{3u}\\
h&=& -\f{g}{u}+H(v).
\end{eqnarray*} 
\end{theorem}

\subsubsection{The case $c=-1$}
\begin{theorem}
For $c=-1$ the integrable deformations of the system  (\ref{fl21-1},\ref{fl21-2}) can be reduced to the following form
\begin{eqnarray*}
\label{intdef1Fr}
u_{t_{(1,1)}}&=&\d_x\alpha_1=\d_x\left[uv+\f{1}{3}u^3+\epsilon^2(a_{16}v_x^2+a_{17}u_xv_x)\right]\\
\label{intdef2Fr}
v_{t_{(1,1)}}&=&\d_x\alpha_2=\d_x\left[\f{1}{2}v^2+\epsilon(a_{21}u_x+a_{22}v_x)+\epsilon^2(a_{23}u_{xx}+a_{24}v_{xx}+a_{25}u_x^2+a_{26}v_x^2+a_{26}uu_xv_x)\right]
\end{eqnarray*}
where
\begin{eqnarray*}
a_{16}&=&-g_{vv}+\f{2f_v^2+h+2ff_{vv}}{2u}-\f{9}{2}\f{g_v}{u^2}+\f{2ff_v+h}{u^3}-\f{g}{u^4}+\f{2}{3}\f{f^2}{u^5}\\
a_{17}&=&ff_{vv}-ug_{vv}+f_v^2-\f{4g_v}{u}+\f{2ff_v}{u^2}+\f{2}{3}\f{f^2}{u^4}\\
a_{21}&=&uf(u,v)\\
a_{22}&=&f(u,v)\\
a_{23}&=&g(u,v)\\
a_{24}&=&h(u,v)\\
a_{25}&=&ug_v-ff_v+\f{3g}{u}-\f{1}{6}\f{f^2}{u^2}\\
a_{26}&=&ug_{vv}-ff_{vv}-f_v^2+\f{6g_v}{u}-\f{11}{3}\f{ff_v}{u^2}+\f{3g}{u^3}-\f{4}{3}\f{f^2}{u^4}\\
%a_{27}&=&-uff_{vv}+u^2g_{vv}-uf_v^2+6g_v-\f{11}{3}\f{ff_v}{u}+\f{3g}{u^2}-\f{4}{3}\f{f^2}{u^3}
\end{eqnarray*}
and
\begin{eqnarray*}
f&=&F(\f{1}{2}u^2+v)u\\
g&=&G(\f{1}{2}u^2+v)u^3+\f{f^2}{3u}\\
h&=&\f{g}{u}+H(v).
\end{eqnarray*} 
\end{theorem}

\subsubsection{The case $c\ne -\f{3}{4},-\f{1}{2},-1$} 
\begin{theorem}
For $c\ne -\f{1}{2},-\f{3}{4},-1$ integrable second order deformations of the system  \eqref{B2-1},\eqref{B2-2} can be reduced to the following form
\begin{eqnarray*}
u_{t_{(1,0)}}=\d_x\alpha_1&=&\d_x\left[\f{1}{2}(-4c-3)u^2+v+\epsilon^2a_{16}(2cuu_x+uu_x-v_x)(2cuu_x+2uu_x-v_x)\right]\\
v_{t_{(1,0)}}=\d_x\alpha_2&=&\d_x\left[\f{1}{3}(-4c^2-6c-2)u^3+\epsilon^2(a_{23}u_{xx}+a_{24}v_{xx}+a_{25}u_x^2+a_{26}v_x^2+a_{27}u_xv_x)\right]
\end{eqnarray*}
where
\begin{eqnarray*}
a_{16}&=&-\f{1}{8}\f{(64c^4+192c^3+252c^2+162c+41)}{(c+1)(2c+1)(4c+3)}\f{\d^2 a_{23}}{\d v^2}+\\
&&\f{1}{16}\f{(64c^4+192c^3+252c^2+162c+41)}{(c+1)^2(2c+1)^2u}\f{\d^2 a_{23}}{\d u\d v}\\
&&-\f{1}{16}\f{(64c^4+192c^3+252c^2+162c+41)}{(c+1)^2(2c+1)^2(4c+3)u^2}\f{\d^2 a_{23}}{\d u^2}+\\
&&-\f{3}{16}\f{(64c^4+192c^3+252c^2+162c+41)}{(c+1)^2(2c+1)^2(4c+3)u^3}\f{\d a_{23}}{\d u}+\\
&&-\f{1}{4}\f{256c^6+1152c^5+2096c^4+1968c^3+960c^2+198c+3}{(c+1)^2(2c+1)^2(4c+3)u^4}a_{23},
\end{eqnarray*}

\begin{eqnarray*}
a_{24}&=&-\f{1}{4}\f{u^3}{4c+3}\f{\d^2 a_{23}}{\d v^2}+\\
&&-\f{1}{8}\f{u^2}{(c+1)(2c+1)}\f{\d^2 a_{23}}{\d u \d v}+\\
&&-\f{1}{8}\f{u}{(c+1)(2c+1)(4c+3)}\f{\d^2 a_{23}}{\d u^2}+\\
&&+\f{3}{8}\f{1}{(2c+1)(4c+3)(c+1)}\f{\d a_{23}}{\d u}+\\
&&-\f{1}{2}\f{4c^2+6c+3}{(c+1)(4c+3)(2c+1)u}a_{23}
\end{eqnarray*}

\begin{eqnarray*}
a_{25}&=&\f{1}{4}(32c^4+96c^3+140c^2+102c+29)u^3\f{\d^2 a_{23}}{\d v^2}+\\
&&+\f{1}{8}\f{u^2(4c+3)(32c^4+96c^3+140c^2+102c+29)}{(2c+1)(c+1)}\f{\d^2 a_{23}}{\d u \d v}+\\
&&+\f{1}{8}\f{u(32c^4+96c^3+140c^2+102c+29)}{(c+1)(2c+1)}\f{\d^2 a_{23}}{\d u^2}+\\
&&+\f{u(2c+1)(c+1)}{4c+3}\f{\d a_{23}}{\d v}+\\
&&-\f{1}{8}\f{(96c^4+288c^3+404c^2+282c+79)}{(2c+1)(c+1)}\f{\d a_{23}}{\d u}+\\
&&-\f{1}{2}\f{128c^6+576c^5+1056c^4+1008c^3+492c^2+90c-5}{(2c+1)(c+1)u}a_{23}
\end{eqnarray*}

\begin{eqnarray*}
a_{26}&=&\f{1}{8}\f{(32c^4+96c^3+140c^2+102c+29)u}{(c+1)(2c+1)}\f{\d^2 a_{23}}{\d v^2}+\\
&&\f{1}{16}\f{(4c+3)(32c^4+96c^3+140c^2+102c+29)}{(c+1)^2(2c+1)^2}\f{\d^2 a_{23}}{\d u \d v}+\\
&&\f{1}{16}\f{(32c^4+96c^3+140c^2+102c+29)}{u(c+1)^2(2c+1)^2}\f{\d^2 a_{23}}{\d u^2}+\\
&&-\f{1}{4}\f{20c^2+30c+11}{(2c+1)(4c+3)(c+1)u}\f{\d a_{23}}{\d v}+\\
&&-\f{1}{16}\f{(8c^2+12c+13)(12c^2+18c+7)}{(c+1)^2(2c+1)^2u^2}\f{\d a_{23}}{\d u}+\\
&&-\f{1}{4}\f{128c^6+576c^5+1056c^4+1008c^3+488c^2+84c-7}{u^3(c+1)^2(2c+1)^2}a_{23},
\end{eqnarray*}

\begin{eqnarray*}
a_{27}&=&-\f{1}{8}\f{(512c^6+2304c^5+4832c^4+5856c^3+4168c^2+1608c+259)u^2}{(2c+1)(4c+3)(c+1)}\f{\d^2 a_{23}}{\d v^2}+\\
&&-\f{1}{16}\f{(512c^6+2304c^5+4832c^4+5856c^3+4168c^2+1608c+259)u}{(c+1)^2(2c+1)^2}\f{\d^2 a_{23}}{\d u \d v}+\\
&&-\f{1}{16}\f{512c^6+2304c^5+4832c^4+5856c^3+4168c^2+1608c+259}{(4c+3)(c+1)^2(2c+1)^2}\f{\d^2 a_{23}}{\d u^2}+\f{\d a_{23}}{\d v}\\
&&\f{1}{16}\f{1536c^6+6912c^5+14432c^4+17376c^3+12296c^2+4728c+761}{u(c+1)^2(2c+1)^2(4c+3)}\f{\d a_{23}}{\d u}+\\
&&+\f{1}{4}\f{2048c^8+12288c^7+31872c^6+46656c^5+41600c^4+22416c^3+6616c^2+744c-35}{(2c+1)^2(c+1)^2(4c+3)u^2}a_{23}.
\end{eqnarray*}

where
$$
a_{23}= u^{4c+4}F_1(-cu^2-u^2+v)+u^{-4c-2}F_2(-cu^2-\f{1}{2}u^2+v).
$$
and $F_1$ and $F_2$ are arbitrary functions of their arguments (which coincide with the Riemann invariants of the dispersionless limit).
\end{theorem}

In the Table 2, we summarize the results obtained for the integrable deformations of the dispersionless hierarchy associated to $B_2$:

\begin{table}[h]
\begin{center}
\begin{tabular}{|p{30mm}|p{30mm}|p{30mm}|}

    \hline
	{\bf Values of $c$ } & {\bf Integrable first order deformations} &
	  {\bf Integrable second order deformations}    \\ 
	\hline
$c\neq -\frac{3}{4},-1,-\frac{1}{2}$  & Miura trivial,\newline no functional parameters & Two functional parameters of a single variable  \\
\hline
$c=-\frac{3}{4}$ \newline(Frobenius)  & Miura trivial, no functional parameters & Two functional parameters of a single variable  \\
\hline
$c=-1$  & One functional parameter of a single variable & Two additional functional parameters of a single variable  \\
\hline
$c=-\frac{1}{2}$  & One functional parameter of a single variable & Two additional functional parameters of a single variable \\
	\hline
\end{tabular}
\end{center}
\caption{\footnotesize Functional parameters for the integrable deformations up to order two, according to the values of $c$}
\label{tab_results}
\end{table} 

\begin{remark}\label{ImportantRemark}
As we have just seen, the integrable deformations at order two in $\epsilon$ depend on two arbitrary functions of a single variable. This result has been obtained applying only Miura transformations of the form \begin{equation}\label{MiuraConsLaw}u^i\mapsto w^i=u^i+\epsilon^2 \partial_x \beta^i,\end{equation} where $\beta^i$ are differential homogenous polynomials of degree $1$. We use these Miura transformations because they preserve the form of conservation laws for the systems under analysis. However, it is not clear a priori if the Miura transformations of the form \eqref{MiuraConsLaw} exhaust the class of Miura transformations that preserve the form of conservation laws for our systems. In order to claim that the two arbitrary functions of a single variable do parameterize the integrable deformations at order two in $\epsilon$, it is necessary to show that they can not be eliminated by any Miura transformation preserving the general form of conservation law. A way to bypass this verification is to compute the Miura invariants of the deformed systems we obtained and verify that precisely the two arbitrary functions of a single variable parameterizing the deformations appear in the Miura invariants. Since the Miura invariants are indeed invariant under any Miura transformation preserving the dispersionless limit, we conclude that the two arbitrary functions can not be eliminated and thus they faithfully parameterize the integrable deformations of order two. This is the procedure that we followed. 
\end{remark}

\subsection{Miura invariants for $B_2$-deformations}
In the case of the principal hierarchies associated with semisimple bi-flat $F$-manifolds we have a distinguished set of Riemann invariants: the 
 canonical coordinates. Evaluating the Miura invariants in this special set of coordinates we clearly obtain functions which are invariants with respect the full group of Miura transformations. Below, we list the Miura invariants for the deformation of $B_2$ corresponding to the four cases $c\neq-\frac{3}{4}, -1, -\frac{1}{2}$ and $c=-\frac{3}{4}$, $c=-1,$ $c=-\frac{1}{2}.$ Taking into account that the canonical coordinates are given by
$$r_1=-(c+\f{1}{2})u^2+v,\,\,r_2=-(c+1)u^2+v,$$
we get: 
\begin{itemize}
\item Case $c\neq -\frac{3}{4}, -1, -\frac{1}{2}$ :
\begin{eqnarray*}
\lambda_1&=&-(2c+1)u-\frac{(u^2)^{-2c}F_2\left(-cu^2-\frac{1}{2}u^2+v\right)}{(2c+1)u^3}\epsilon^2+\mathcal{O}(\epsilon^3)=\\
&=&-(2c+1)\sqrt{2r_1-2r_2}-\f{(2r_1-2r_2)^{-2c-\f{3}{2}}F_2(r_1)}{2c+1}\epsilon^2+\mathcal{O}(\epsilon^3)\\
\lambda_2&=&-2u(c+1)-\frac{1}{2}\frac{u^3F_1\left(-cu^2-u^2+v\right)(u^2)^{2c}}{c+1}\epsilon^2+\mathcal{O}(\epsilon^3)=\\
&=&-2\sqrt{2r_1-2r_2}(c+1)-\f{F_1(r_2)(2r_1-2r_2)^{2c+\f{3}{2}}}{2c+2}\epsilon^2+\mathcal{O}(\epsilon^3)
\end{eqnarray*}
\item Case $c=-\frac{3}{4}$
\begin{eqnarray*}
\lambda_1&=&\frac{1}{2}u+\frac{1}{2}\frac{u\partial_v F+2\partial_uF}{u}\epsilon^2+\mathcal{O}(\epsilon^3),\\ 
\lambda_2&=&-\frac{1}{2}u+\frac{1}{2}\frac{u\partial_{v}F-2\partial_u F}{u}\epsilon^2+\mathcal{O}(\epsilon^3),
\end{eqnarray*}
where  $F=F_1(\f{1}{4}u^2+v)+F_2(-\f{1}{4}u^2+v)$. In canonical coordinates:
\begin{eqnarray*}
\lambda_1&=&\f{1}{2}\sqrt{2r_1-2r_2}+F_1'(r_1)\epsilon^2+\mathcal{O}(\epsilon^3),\\
\lambda_2&=&-\f{1}{2}\sqrt{2r_1-2r_2}+F_2'(r_2)\epsilon^2+\mathcal{O}(\epsilon^3),
\end{eqnarray*}
\item Case $c=-1$:
\begin{eqnarray*}
\lambda_1&=&u^2+v+F_1\left(\frac{1}{2}u^2+v \right)u\epsilon+\left(F_2\left(\frac{1}{2}u^2+v \right)u^2+\frac{1}{3}F_1\left(\frac{1}{2}u^2+v \right)^2\right)\epsilon^2+\mathcal{O}(\epsilon^3) =\\
&=&2r_1-r_2+F_1(r_1)\sqrt{2r_1-2r_2}\epsilon+\left(2F_2(r_1)(r_1-r_2)+\f{1}{3}F_1(r_1)^2\right)\epsilon^2+\mathcal{O}(\epsilon^3)\\
\lambda_2&=&v+F_3(v)\epsilon^2+\mathcal{O}(\epsilon^3)=r_2+F_3(r_2)\epsilon^2+\mathcal{O}(\epsilon^3).
\end{eqnarray*}
\item Case $c=-\frac{1}{2}$:
\begin{eqnarray*}
\lambda_1&=&v+F_3(v)\epsilon^2+\mathcal{O}(\epsilon^3)=r_1+F_3(r_1)\epsilon^2+\mathcal{O}(\epsilon^3),\\
\lambda_2&=& v-u^2+F_1\left(v-\frac{1}{2}u^2\right)u\,\epsilon-\left(F_2\left(v-\frac{1}{2}u^2\right)u^2+\frac{1}{3}F_1\left(v-\frac{1}{2}u^2\right)^2\right)\epsilon^2+\mathcal{O}(\epsilon^3)=\\
&=&-r_1+2r_2+F_1(r_2)\sqrt{2r_1-2r_2}\,\epsilon+\left(2F_2(r_2)(r_2-r_1)-\f{1}{3}F_1(r_2)^2\right)\epsilon^2+\mathcal{O}(\epsilon^3).
\end{eqnarray*}
\end{itemize}

\section{The case $I_2(m)$}\label{sec6}
The case of $I_2(m)$  can be treated in a completely analogous way  (this is not surprising since $I_2(4)\sim B_2$). We skip details of the computations, which are summarized in the Table 3. For any $m$ we have a special
 values of the parameter ($c=0$) corresponding to the Dubrovin-Saito Frobenius manifold structure and two special values ($c=\pm 2$) such that one of the primary flows is degenerate. In this case, the first order non-trivial deformations depend on a functional parameter and second order deformations contain two additional functional parameters. In all the remaining cases, the first order deformations are trivial and the second order deformations depend on two functional parameters. To simplify the formulas we have substituted $m$ with $2m$ everywhere.
 
\begin{table}[h]
\begin{center}
\begin{tabular}{|p{30mm}|p{30mm}|p{30mm}|}

    \hline
	{\bf Values of $c$ } & {\bf Integrable first order deformations} &
	  {\bf Integrable second order deformations}    \\ 
	\hline
$c\neq 0,\pm 2$  & Miura trivial,\newline no functional parameters & Two functional parameters of a single variable  \\
\hline
$c=0$ \newline(Frobenius)  & Miura trivial, no functional parameters & Two functional parameters of a single variable  \\
\hline
$c=\pm 2$  & One functional parameter of a single variable & Two additional functional parameters of a single variable  \\
\hline
\end{tabular}
\end{center}
\caption{\footnotesize Functional parameters for the integrable deformations up to order two, according to the values of $c$}
%\label{tab_results}
\end{table} 

\subsubsection{The case $c=0$}
\begin{theorem}
For $c=0$ the integrable second order deformations of the system  (\ref{B2-1},\ref{B2-2}) can be reduced to the following form
\begin{eqnarray}
\label{intdef1Fr}
u_{t_{(1,1)}}&=&\d_x\left[v+\epsilon^2(a_{15}u_x^2+a_{16}v_x^2+a_{17} u_xv_x)\right]\\
\label{intdef2Fr}
v_{t_{(1,1)}}&=&\d_x\left[\f{4m^2}{2m-1}u^{2m-1}+\epsilon^2(a_{23}u_{xx}+a_{24}v_{xx}+a_{25}u_x^2-\f{u^{-2m+2}a_{25}}{4m^2}v_x^2)\right]
\end{eqnarray}
where
\begin{eqnarray*}
a_{15}&=&\f{3}{8}F_{vv}u^{2m-3}+\f{1}{32}\f{m-2}{m(m-1)}\f{F_u}{u^{2}},\\
a_{16}&=&-\f{1}{32 m^2}\f{F_{vv}}{u}-\f{1}{128}\f{m-2}{m^3(m-1)}\f{F_u}{u^{2m}},\\
a_{17}&=&\f{1}{16 m^2}\f{F_{uv}}{u},\\
a_{23}&=&-\f{1}{2}F_vu^{2m-3},\\
a_{24}&=&-\f{1}{8m^2}\f{F_u}{u},\\
a_{25}&=&-\f{1}{4}F_{uv}u^{2m-3}+\f{1}{8}\f{2m-3}{m-1}F_v u^{2m-4}.
\end{eqnarray*}
and $F$ is the sum of two arbitrary functions of the Riemann invariants: 
$$F=F_1(v+2u^m)+F_2(v-2u^m).$$
\end{theorem}
 
 \subsubsection{The case $c=\pm 2$}
\begin{theorem}
For $c=\pm 2$ the integrable deformations of the system (\ref{fl21-1},\ref{fl21-2}) can be reduced to the following form
\begin{eqnarray*}
\label{intdef1Fr}
u_{t_{(1,1)}}&=&\d_x\left[\mp\f{4mu^{m+1}}{m+1}+uv+\epsilon^2(a_{16}v_x^2\mp 4mu^{m-1}a_{16}u_xv_x)\right]\\
\label{intdef2Fr}
v_{t_{(1,1)}}&=&\d_x\left[\f{1}{2}v^2+\epsilon \left(\mp 4m u^{2m-2}u_x+u^{m-1}v_x\right)F_1+\epsilon^2(a_{23}u_{xx}+a_{24}v_{xx}+a_{25}u_x^2+\epsilon^2a_{26}v_x^2+a_{27}u_xv_x)\right]
\end{eqnarray*}
where
\begin{eqnarray*}
a_{16}&=&-\f{1}{48}\f{m-4}{m^3}\f{FF_v}{u}-\f{1}{192}\f{m-4}{m^4}\f{F^2}{u^{m+1}}-\f{1}{64}\f{(m-3)(m-1)(m+1)}{m^2(m-2)}\f{f}{u^{2m-1}}\\
a_{23}&=&\f{2}{3}\f{m-1}{m}u^{2m-3}F^2+u^{3m-3}G\\
a_{24}&=&-\f{1}{6}\f{m-1}{m^2}u^{m-2}F^2-\f{1}{4m}u^{2m-2}G-\f{1}{4}\f{m+1}{m}f\\
a_{25}&=&3(m-1)u^{3m-4}G-4mu^{4m-4}G_v+\f{1}{6}\f{8m^2-17m+12}{m}u^{2m-4}F^2-\f{4}{3}(m-4)u^{3m-4}FF_v\\
a_{26}&=&-\f{1}{12}\f{m-4}{m^2}u^{m-2}FF_v-\f{1}{4m}u^{2m-2}G_v-\f{1}{96}\f{m-4}{m^3}\f{F^2}{u^2}+\f{1}{16}\f{m-1}{m^2}u^{m-2}G+\\
&&-\f{1}{8}\f{(m+1)(m-3)}{m(m-2)}f_v-\f{1}{16}\f{(m+1)(m-3)(m-1)^2}{m^2(m-2)}\f{f}{u^{m}}\\
a_{27}&=&\f{2}{3}\f{m-4}{m}u^{2m-3}FF_v+2u^{3m-3}G_v-\f{1}{12}\f{2m^2-5m+6}{m^2}u^{m-3}F^2+\\
&&-\f{3}{4}(m-1)u^{2m-3}G+\f{1}{4}\f{(m-1)(m-3)(m+1)}{(m-2)}\f{f}{u},
\end{eqnarray*}
and
\begin{eqnarray*}
F&=&F(v\mp 4u^m)\\
G&=&G(v\mp 4u^m)\\
f&=&f(v).
\end{eqnarray*} 
\end{theorem}

\subsubsection{The case $c\ne 0,\pm 2$} 
\begin{theorem}
For $c\ne 0,\pm 2$ integrable second order deformations of the system  \eqref{B2-1},\eqref{B2-2} can be reduced to the following form
\begin{eqnarray*}
u_{t_{(1,0)}}&=&\d_x\left[-2u^mc+v+\epsilon^2a_{16}(u^{2m-2}m^2(c^2-4)u_x^2-2cmu^{m-1}u_xv_x+v_x^2)\right]\\
v_{t_{(1,0)}}&=&\d_x\left[-\f{m^2}{2m-1}(c^2-4)u^{2m-1}+\epsilon^2(a_{23}u_{xx}+a_{24}v_{xx}+a_{25}u_x^2+a_{26}v_x^2+a_{27}u_xv_x)\right]
\end{eqnarray*}
where
\begin{footnotesize}
\begin{eqnarray*}
a_{16}&=&\f{1}{16}\f{c^4m^2-2c^4m+c^4+12c^2m^2-4c^2m-4c^2+16m}{m^2(m-1)^2(c-2)(c+2)cu^{m-2}}\f{\d^2 a_{23}}{\d v^2}+\\
&&\f{1}{8}\f{c^4m^2-2c^4m+c^4+12c^2m^2-4c^2m-4c^2+16m}{m^3(m-1)^2(c-2)^2(c+2)^2u^{2m-3}}\f{\d^2 a_{23}}{\d u\d v}+\\
&&\f{1}{16}\f{c^4m^2-2c^4m+c^4+12c^2m^2-4c^2m-4c^2+16m}{m^4(m-1)^2(c-2)^2(c+2)^2cu^{3m-4}}\f{\d^2 a_{23}}{\d u^2}+\\
&&-\f{1}{8}\f{(m-2)(c^4m^2-2c^4m+c^4+12c^2m^2-4c^2m-4c^2+16m)}{m^3(m-1)^2(c-2)^2(c+2)^2u^{2m-2}}\f{\d a_{23}}{\d v}+\\
&&-\f{1}{16}\f{(5m-7)(c^4m^2-2c^4m+c^4+12c^2m^2-4c^2m-4c^2+16m)}{m^4(m-1)^2(c-2)^2(c+2)^2cu^{3m-3}}\f{\d a_{23}}{\d u}+\\
&&\f{c^2(19m^3-32m^2+3m+9)+24m^2-36m}{4(m-1)m^4c(c-2)^2(c+2)^2u^{3m-2}}a_{23}+\\
&&-\f{c^6(m-1)^2-4c^4(6m^2-15m+10)}{64m^4c(c-2)^2(c+2)^2u^{3m-2}}a_{23},
\end{eqnarray*}

\begin{eqnarray*}
a_{24}&=&-\f{4u^{m+1}}{(m-1)c}\f{\d^2 a_{23}}{\d v^2}-\f{8u^2}{m(m-1)(c-2)(c+2)}\f{\d^2 a_{23}}{\d u \d v}+\\
&&-\f{4u^{-m+3}}{m^2(m-1)c(c-2)(c+2)}\f{\d^2 a_{23}}{\d u^2}+\f{8u(m-2)}{m(m-1)(c-2)(c+2)}\f{\d a_{23}}{\d v}+\\
&&\f{4(5 m-7)u^{-m+2}}{m^2(m-1)c(c-2)(c+2)}\f{\d a_{23}}{\d u}-\f{(c^2+24m-36)u^{-m+1}}{c(c-2)(c+2)m^2}a_{23}
\end{eqnarray*}

\begin{eqnarray*}
a_{25}&=&\f{m}{16}\f{((m-1)^2c^4+(24m^2-12m-8)c^2+16m^2+16m+16)u^{2m-1}}{(m-1)^2}\f{\d^2 a_{23}}{\d v^2}+\\
&&\f{c}{8}\f{((m-1)^2c^4+(24m^2-12m-8)c^2+16m^2+16m+16)u^{m}}{(m-1)^2(c-2)(c+2)}\f{\d^2 a_{23}}{\d u \d v}+\\
&&\f{1}{16}\f{((m-1)^2c^4+(24m^2-12m-8)c^2+16m^2+16m+16)u}{m(m-1)^2(c-2)(c+2)}\f{\d^2 a_{23}}{\d u^2}+\\
&&-\f{1}{8}\f{((m-2)c^6-32m)u^{m-1}}{c(c-2)(c+2)}\f{\d a_{23}}{\d v}-\f{(5m-7)c^4}{16m(c-2)(c+2)}\f{\d a_{23}}{\d u}+\\
&&-\f{1}{8}\f{((22m^3-56m^2+14m+16)c^4+(32m^3-48m^2-32)c^2)u^{m-1}}{(m-1)^2c(c-2)(c+2)}\f{\d a_{23}}{\d v}+\\
&&-\f{((104m^3-196m^2+28m+56)c^2+144m^3-160m^2+32m-112)}{16m(m-1)^2(c-2)(c+2)}\f{\d a_{23}}{\d u}+\\
&&-\f{(m-1)^3c^6-(24m^3-88m^2+108m-44)c^4-(560m^3-976m^2+112m+304)c^2}{64(m-1)m(c-2)(c+2)u}a_{23}+\\
&&+\f{8m^3-8m^2+m-9}{(m-1)m(c-2)(c+2)u}a_{23},\\
\end{eqnarray*}

\begin{eqnarray*}
a_{26}&=&\f{((m-1)^2c^4+4(6m^2-3m-2)c^2+16(m^2+1m+1))u}{16m(m-1)^2(c-2)(c+2)}\f{\d^2 a_{23}}{\d v^2}+\\
&&\f{c((m-1)^2c^4+4(6m^2-3m-2)c^2+16(m^2+1m+1))}{8m^2(m-1)^2(c-2)^2(c+2)^2u^{m-2}}\f{\d^2 a_{23}}{\d u \d v}+\\
&&\f{((m-1)^2c^4+4(6m^2-3m-2)c^2+16(m^2+1m+1))}{16m^3(m-1)^2(c-2)^2(c+2)^2u^{2m-3}}\f{\d^2 a_{23}}{\d u^2}+\\
&&-\f{(m-2)c^6+32m}{8m^2c(c-2)^2(c+2)^2u^{m-1}}\f{\d a_{23}}{\d v}-\f{(5m-7)c^4}{16m^3(c-2)^2(c+2)^2u^{2m-2}}\f{\d a_{23}}{\d u}+\\
&&-\f{(34m^3-80m^2+26m+16)c^4+(-32m^3+80m^2-64m-32)c^2}{8m^2(m-1)^2c(c-2)^2(c+2)^2u^{m-1}}\f{\d a_{23}}{\d v}+\\
&&-\f{((128m^3-244m^2+52m+56)c^2+48m^3+32m^2-64m-112}{16m^3(m-1)^2(c-2)^2(c+2)^2u^{2m-2}}\f{\d a_{23}}{\d u}+\\
&&-\f{(-656m^3+1200m^2-240m-304)c^2-128m^3-384m^2+448m+576}{64m^3(m-1)(c-2)^2(c+2)^2u^{2m-1}}a_{23},\\
&&-\f{(m-1)^2c^6-4(6m^2-16m+11)c^4}{64m^3(c-2)^2(c+2)^2u^{2m-1}}a_{23},\\
\end{eqnarray*}

\begin{eqnarray*}
a_{27}&=&-\f{((m-1)^2c^6+4(6m^2-3m-2)c^4+16(m^2-m+3)c^2+128(m-1))u^m}{8(m-1)^2c(c-2)(c+2)}\f{\d^2 a_{23}}{\d v^2}+\\
&&-\f{((m-1)^2c^6+4(6m^2-3m-2)c^4+16(m^2-m+3)c^2+128(m-1))u}{4m(m-1)^2(c-2)^2(c+2)^2}\f{\d^2 a_{23}}{\d u \d v}+\\
&&-\f{((m-1)^2c^6+4(6m^2-3m-2)c^4+16(m^2-m+3)c^2+128(m-1))}{8m^2(m-1)^2c(c-2)^2(c+2)^2u^{m-2}}\f{\d^2 a_{23}}{\d u^2}+\\
&&\f{(5m-7)c^6}{8m^2c(c-2)^2(c+2)^2u^{m-1}}\f{\d a_{23}}{\d u}+\f{(m-2)c^6}{4m(c-2)^2(c+2)^2}\f{\d a_{23}}{\d v}+\\
&&\f{(28m^3-68m^2+20m+16)c^4+(-16m^3+16m^2+48m-96)c^2+64m^3-320m+256}{4m(c-2)^2(c+2)^2(m-1)^2}\f{\d a_{23}}{\d v}+\\
&&\f{(29m^3-55m^2+10m+14)c^4+4(7m^3-16m^2+24m-21)c^2-16(m^3-12m^2-25m+14)}{2m^2(m-1)^2c(c-2)^2(c+2)^2u^{m-1}}\f{\d a_{23}}{\d u}+\\
&&\f{(-768m^3+1920m^2-2496m+1856)c^2+512m^3-4352m^2+8448m-4608}{32m^2(m-1)c(c-2)^2(c+2)^2u^{m}}a_{23},\\
&&\f{(m-1)^2c^8-4(6m^2-16m+11)c^6-16(33m^2-23m-21)c^2}{32m^2c(c-2)^2(c+2)^2u^{m}}a_{23},
\end{eqnarray*}
\end{footnotesize}
\begin{eqnarray*}
a_{23}&=&u^{\f{1}{2}c(m-1)+2m-3}F_1(v-(c+2)u^m)+u^{-\f{1}{2}c(m-1)+2m-3}F_2(v-(c-2)u^m).
\end{eqnarray*}
and $F_1$ and $F_2$ are arbitrary functions of their arguments (which coincide with the Riemann invariants of the dispersionless limit).
\end{theorem}

\section{Conclusions}\label{sec7}
In this paper following the path already traced in the scalar case \cite{ALM,ALM2}
 we have generalized Dubrovin-Zhang approach to the study
 of integrable hierarchies of systems of conservation laws. 
In this framework Frobenius manifolds are replaced by flat
 and bi-flat $F$-manifolds, Poisson bracket between hamiltonian functionals
 are replaced by Poisson brackets between currents
 and central invariants are replaced by Miura invariants. Due to the
 complexity of computations we have considered 2-component systems
 and only first order and second order deformations. This is enough to
 see some new interesting phenomena like the appearance of new functional
 parameters in some special cases.  However it is clear that important question like the existence of eventual 
 obstructions in the deformation procedure require
 new ideas and new tecniques. Unfortunately a powerful instrument
 like bi-Hamiltonian cohomology is not available in this setting.
 \newline
 \newline
Other important issues are:
 \begin{itemize}
\item To find examples of truncated integrable systems of conservation laws.
  These might be obtained selecting special values for
 the functional parameters on which the deformation depends. 
\item To study the (possibly universal) behaviour of solutions
 near the point of gradient catastrophe and how it changes varying 
 the parameter. It is reasonable to expect substantial differences between
 the Hamiltonian and the non-Hamiltonian cases.     
\item To study the non-semisimple case.  The deformations procedure can be applied without significant changes also to the principal hierachy of a non semisimple (bi)-flat $F$-manifold. Some results are already available in the bi-Hamiltonian case \cite{DVS}.
\end{itemize}

\section*{Ackowledgements} 
This research was partially supported by GNFM through the 2017 Visitors Program and by the Department  of Mathematics and Applications of the Universisty
 of Milano-Bicocca through funds for visitors. AA also thanks the Department of Mathematics and Applications of the University of Milano-Bicocca for its warm
hospitality and stimulating environment. PL thanks Paolo Rossi for useful discussions.

\end{document}